\documentclass[twocolumn]{aastex62}

\usepackage{graphicx,amsmath,amssymb,amstext,natbib}

\usepackage{epsf}
\usepackage{epstopdf}
\usepackage{xspace}
\usepackage{color}

\newcommand{\msun}{{M_{\odot}}}
\newcommand{\mstar}{{M_{\ast}}}
\newcommand{\ser}{S\'ersic }

\shorttitle{Stellar Mass profiles of Quiescent Galaxies in Different Environments at $z\sim0$}
\shortauthors{Mosleh et al.}

\journalinfo{Accepted for publication in ApJ}

\begin{document}

\title{Stellar Mass profiles of Quiescent Galaxies in Different Environments at $z\sim0$}

\correspondingauthor{Moein Mosleh}
\email{moein.mosleh@shirazu.ac.ir}

\author[0000-0002-4111-2266]{Moein Mosleh}
\affiliation{Biruni Observatory, Shiraz University, Shiraz 71454, Iran}
\affiliation{Physics Department, Shiraz University, Shiraz 71454, Iran}

\author[0000-0003-0126-8554]{Saeed Tavasoli}
\affiliation{Faculty of Physics, Kharazmi University, Mofateh Ave., Tehran, Iran}

\author[0000-0002-8224-4505]{Sandro Tacchella}
\affiliation{Harvard-Smithsonian Center for Astrophysics, 60 Garden St, Cambridge, MA 02138, USA}

\begin{abstract}

We present the stellar mass profiles of 147 \textit{isolated quiescent} galaxies in very low-density environments (i.e., void regions) in the local Universe ($0.01<z<0.06$) from the Sloan Digital Sky Survey. These galaxies have stellar masses between $ 9.8\lesssim\log(\mstar/\msun)\lesssim11.2$ and they represent $\sim15\%$ of the whole galaxy population in the void regions down to $M_{r} = -19$. We do not find any isolated quiescent galaxies with $\log(\mstar/\msun)\gtrsim11.2$. We compare the stellar mass profiles of these isolated quiescent galaxies with the profiles of stellar mass-matched samples of the quiescent galaxies in group and cluster environments. We find that, at fixed mass, quiescent galaxies in voids have similar central ($1$ kpc) mass densities ($\Sigma_1$) and central velocity dispersions ($\sigma_1$) compared to their counterparts in groups and clusters. We show that quiescent galaxies in voids have at most $10-25\%$ smaller half-mass (and half-light) sizes compared to quiescent galaxies in groups and clusters. We conclude that for the intermediate stellar mass range of $10^{10}-10^{11}\msun$ in the local Universe, environmental mechanisms have no significant additional effect on the mass profiles of the quiescent galaxies. \\

\end{abstract}

 
\section{Introduction}
\label{introduction}

The mechanisms driving the stellar mass assembly within galaxies is not well-understood. Although the higher fraction of early-type morphologies in the higher local densities compared to the low-density environments, known as the morphology-density relation \citep{dressler1980, postman1984}, has already been observed, it is not clear how much the environments contribute to the transformation of disk galaxies into the spheroids, at least for intermediate stellar mass systems \citep[see e.g.,][]{Carollo2016}. The mixture of different processes, such as disk instabilities, minor/major mergers, formation/migration of giant clumps of gas might contribute to these morphological transformations \citep[e.g.,][]{kormendy2004, bell2006, khochfar2006, genzel2008,  Elmegreen2008, hilz2013, Bournaud2016}. However, the fractional contribution of each mechanism and their dependence on galaxy stellar mass and environment needs to be investigated. For instance, if the frequency of (minor/major) mergers is higher in denser environments, then it is expected that these processes affect the stellar mass distributions within galaxies (e.g., in the outer parts of galaxies) in the high-density regions compared to the low-density regions \citep{fakhouri2009, lin2010, shankar2013, shankar2014}.

Scaling relations are promising tools for exploring the contribution of different physical mechanisms to the formation of galaxies structures, especially, if we compare them for different environments. The relation between the stellar mass and size of the galaxies (in particular for quiescent/early-type galaxies) is one of the important ones \citep{shen2003}. As mentioned above, in  high-density regions, galaxies can experience more dissipationless dry mergers or be affected by the galaxy harassment or tidal stripping \citep{Spindler2017}. If the evolution of galaxies' structures and hence sizes accelerated by the mergers (as predicted from the hierarchical galaxy formation) or affected by aforementioned effects in the high-density regions, then one should expect the dependence of the mass-size relation on environment. Many studies have already examined this for both low and high redshift galaxies \citep[e.g.,][]{galletta2006, maltby2010, Nair2010, cooper2012, lani2013, lorenzo2013, huertas2013, shankar2014, cebria2014, kelkar2015, allen2015, Zhao2015, lacerna2016,  yoon2017, saracco2017}. However, there is no consensus among different studies, in particular at high redshifts. Some authors found no differences between the size of galaxies at fixed stellar masses and some found that quiescent/early-types are larger in the local over-dense regions. This discrepancy is partially suggested by some authors  \citep[e.g.,][]{saracco2017} to be due to the biases introduced in the sample selection methods (hence the different mixture of the galaxies and different stellar mass ranges) and/or the definition of environment \citep{bassett2013}.

As mentioned above, different criteria for defining environment can be one of the drivers of the discrepancy among many studies. Various definition of the high-density regions, such as using the projected nearest neighbor surface density or the X-ray detected clusters in comparison to the field or average galaxy populations as the low-density environments, makes it hard for a true comparison of the results among different studies \citep[e.g., comparing these studies:][]{cooper2012, lacerna2016, huertas2013}. Thanks to the redshift surveys such as the Sloan Digital Sky Survey (SDSS), we can trace the distributions of the galaxies over a large cosmological volume in the local Universe. The distributions depict structures such as cosmic filaments, walls, over-dense regions of the galaxies (groups/clusters) and large volumes in the space which are empty of the galaxies and are known as voids \citep{Einasto1980, kirshner1981}. Despite their emptiness, void environments hold a large total volume of the Universe and hence suitable regions for finding very isolated galaxies \citep{Aragon2010, vandeweygaert2011, tavasoli2013}. Selecting galaxies from these vast environments compared to the extremely high-density regions (i.e., void galaxies compared to cluster galaxies), should magnify any variation between galaxy parameters at different environments; i.e., void galaxies are the best probe for distinguishing the environmental effects on the galaxies \citep[e.g.,][]{Grogin1999a, Grogin2000, Colbert2001, Denicolo2005a, Denicolo2005b, Collobert2006, tavasoli2015, mooran2015, fraser-mckelvie2016, beygu2017, Kuutma2017}.  In this work, for the first time, we used a sample of isolated void galaxies compared to the cluster galaxies, to study the effect of the environment on the mass-size relation.

It is also important to emphasize that the existence of color gradients \citep{franx1989, peletier1990, labarbera2005} and hence mass to light ratio ($\mstar/L$) radial variations need also to be considered. The buildup of the central densities (or bulges) and the outer parts of the galaxies can be better assessed by means of the stellar mass density profiles after correcting for the age/metallicity degeneracy \citep{cheung2012}. Therefore, studying the stellar mass profiles of the galaxies is better suited for investigating the role of environment on the mass assembly of the galaxies compared to the light profiles \citep[e.g.,][]{fang2013, tacchella2015, Carollo2016, Chan2016, mosleh2017, Barro2017}. This study benefits of using mass-weighted profiles of galaxies compared to the previous works on the mass-size relation, and hence reducing systematic errors due to any environmental dependence on the color-gradients of galaxies.

On the other hand, the responsible mechanisms for the cessation of the star-formation activity of galaxies and transferring galaxies from the blue cloud into the red sequence (i.e., ``quenching'') is yet to be established. As shown by \cite{Penglilly2010}, two distinct modes of the quenching are in general acting, i.e., ``mass'' and ``environment'', for classifying the quenching mechanisms, especially for the central and satellite galaxies. By minimizing the effect of the environment, one can examine different mass quenching mechanisms such as the halo quenching \citep{brinboim2003,keres2005,dekel2006}, active galactic nucleus (AGN) feedback  or the existence of a supermassive black hole \citep[][]{ Dimatteo2005, spingel2005, croton2006} and the total stellar masses.  The correlation between different properties of the galaxies is helpful for evaluating different processes related to the mass quenching \citep[although the correlations do not imply causal connection, see e.g.,][]{Lilly2016}. Among them, the properties related to the central regions of galaxies are promising tools \citep{allen2006, Driver2006, schiminovich2007, bell2008, mendez2011, lang2014}. As shown recently  \citep{cheung2012, fang2013, whitaker2017, Barro2017, mosleh2017, tacchella2017}, the central mass density of galaxies has a tight correlation with the color or the star-formation activity of galaxies at low and high redshifts. The central velocity dispersion is also shown to be a good predictor of quenching \citep[e.g.,][]{wake2012}. For quiescent galaxies, both of these properties are tightly correlated with the total stellar mass of galaxies. Comparing these scaling relations for the quiescent galaxies in different environments can help in assessing the mass (halo) quenching mechanisms in details. Therefore \textit{isolated quenched void} galaxies are in particular interesting, as these galaxies are expected to have suffered less from environmental processes acting during quenching.  
 
In this paper, by finding  quiescent galaxies in the extremely low (isolated ones in the voids), intermediate (groups) and high-density (clusters) environments, we aim to compare the stellar mass distributions within these galaxies (which is related to the morphological transformation) and the scaling relations such as the correlation between the central density/velocity-dispersion with the total stellar of quiescent galaxies in different environments.   

The structure of this paper is as follows. In Section 2, we describe how the samples are selected and in Section 3, the methodology for deriving light and stellar mass profiles are described. The results are presented and discussed in Section 4, and 5. We adopt the following cosmological parameters for this work: $\Omega_{m}$ = 0.3, $\Omega_{\Lambda}$ = 0.7 and $H_{0} = 70$ $km$ $s^{-1}$ $Mpc^{-1}$.\\

\begin{figure}
\includegraphics[width=0.5\textwidth]{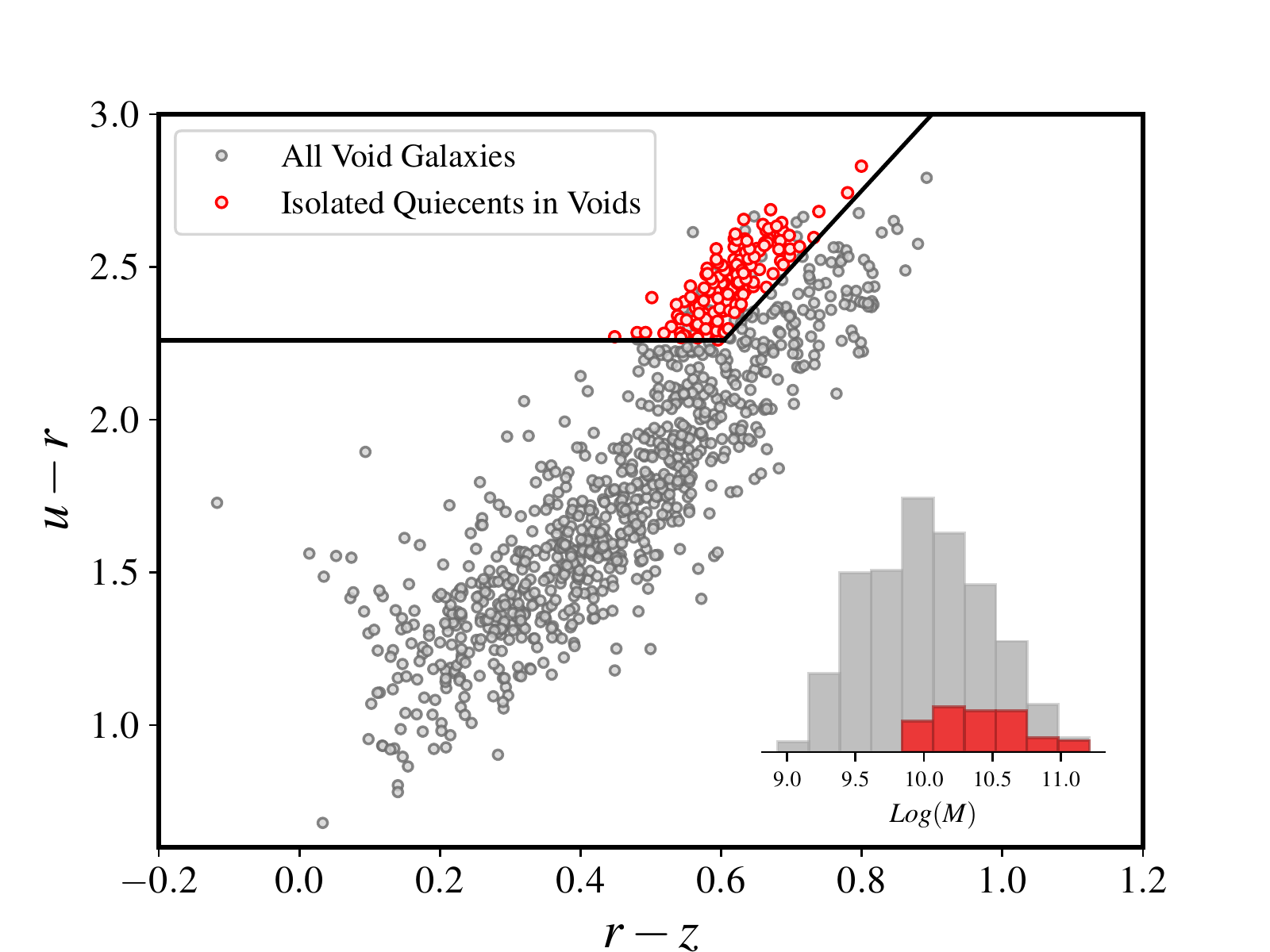}
\vspace{5. mm}
\includegraphics[width=0.5\textwidth]{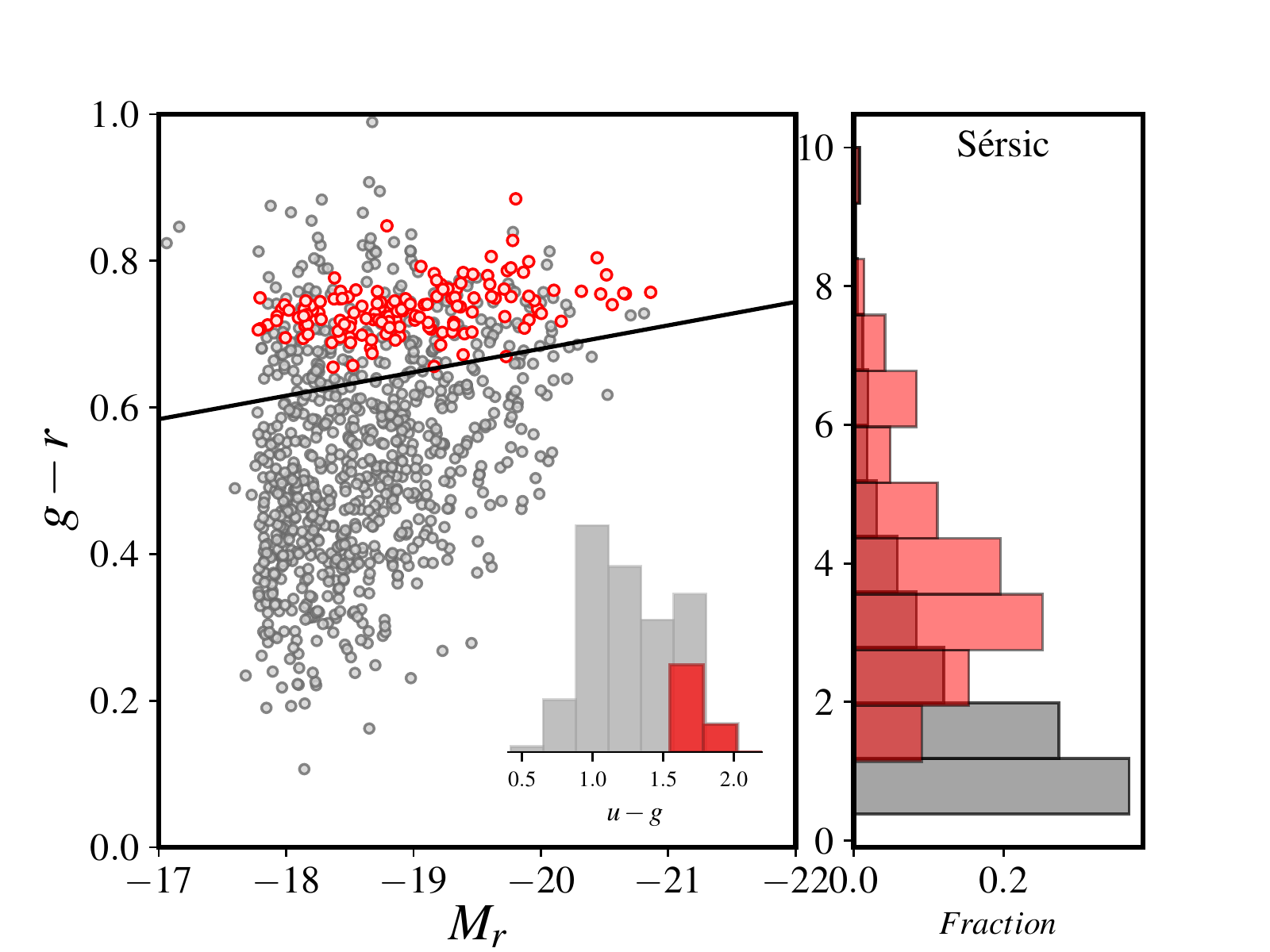}
\caption{\textit{Top Panel}: The criteria used for separating the quiescent galaxies from the star-forming ones based on the $u-r$ versus $r-z$ color-color diagram \citep{holden2012}. The gray symbols are all void galaxies, and the red ones are the isolated quiescent void galaxies in this study. The inset shows the stellar mass distributions of all void galaxies (gray histogram) and the isolated quiescent sample (red histogram).  \textit{Bottom Panel}: The locations of isolated quiescent galaxies in the $g-r$ color-magnitude diagram. The solid line criteria used in this figure is defined as $ g -r  > 0.68 - 0.032\times (M_{r}+20)$ \citep[see][]{mosleh2013}. The inset histogram depicts $u-g$ distributions and the histograms in the right bottom panel show the distributions of the \ser indices of the samples.} 
\label{fig1}
\end{figure}

\section{DATA \& SAMPLE}
\label{data}

Our sample of the quiescent galaxies in the void regions is extracted from a catalog of void galaxies by \cite{tavasoli2015} based on the Sloan Digital Sky Survey Data Release 10 (SDSS DR10) \citep{Ahn2014}. In this catalog, there are 1014 galaxies down to a limiting magnitude of $M_{r} = -19.$ and at the redshift range of $0.010 \leq z \leq 0.055$. The sample is also complete down to $10^{10}\msun$. These galaxies reside in the 167 void regions which are selected by applying a void finder algorithm introduced by \cite{Aikio1998} \citep[see][for a review on different void algorithms]{colberg2008}. As described in \cite{tavasoli2013}, the final void finding algorithm excludes spurious voids with radii $\leq 7$ Mpc. Hence, the void regions are extremely large empty volumes in the local Universe and have large background density contrasts. This ensures that galaxies in this sample reside in a totally different environments than galaxies in walls (including filaments, groups, and clusters) of the cosmic web. We caveat that due to the structures within the voids \citep{Aragon2013} and the redshift space distortion effects, there is not a unique definition for the voids.

The stellar masses and spectroscopic properties of the galaxies are taken from the Max-Planck-Institute for Astrophysics (MPA)-Johns Hopkins University (JHU) SDSS DR7 catalog \citep{kaufmann2003a, brinchmann2004, salim2007} and the galaxy colors are from the New York University Value-Added Galaxy Catalog \citep[NYU-VAGC;][]{Blanton2005}. For our 1014 void galaxies, 980 objects have measured stellar masses in the MPA-JHU catalog. In order to separate quiescent from star-forming galaxies, we follow a similar $u-r$ and $r-z$ color-color criteria defined by \cite{holden2012}:

\begin{equation}
(u -r)_{0} > 2.26
\end{equation}
\begin{equation}
(u -r)_{0} > 0.76+2.5(r-z)_{0}
\end{equation}

This is similar to the $UVJ$ selection (i.e., $U-V$ versus $V-J$ rest-frame colors) used at high-$z$ \citep{williams2009a}, which separates unobscured and dusty star-forming galaxies from quiescent ones (see Figure \ref{fig1}).  Based on these criteria, we select 174 out of the 980 void galaxies to be quiescent ($\sim18\%$).  In order to understand better the influence of the environments on the mass profile of the quiescent galaxies in the void regions, we then focus on isolated quiescent galaxies in these environments. Therefore, we cross-matched our void galaxies with a catalog of groups provided in \cite{Tempel2014} and removed any galaxies which have more than one companion. This reduces the number of the isolated quiescent galaxies to 147 in the void region (15\% of the void galaxies in our catalog).   We find that this additional selection does not have any significant effect on the stellar-mass distribution. The top panel of Figure \ref{fig1} shows the region where these 147 isolated quiescent void galaxies are separated from the rest of the void galaxies. The inset shows the distribution of the stellar masses of the isolated quiescent void galaxies (red histogram) compared to the general mass distribution of the void galaxies. The isolated quiescent void galaxies have the stellar masses between $9.8 \lesssim \log(\mstar/\msun) \lesssim 11.2$ and 136 objects ($92\%$) of this sample have the stellar masses $\log(\mstar/\msun) \gtrsim 10.$

\begin{figure}
\centering
\includegraphics[width=0.4\textwidth]{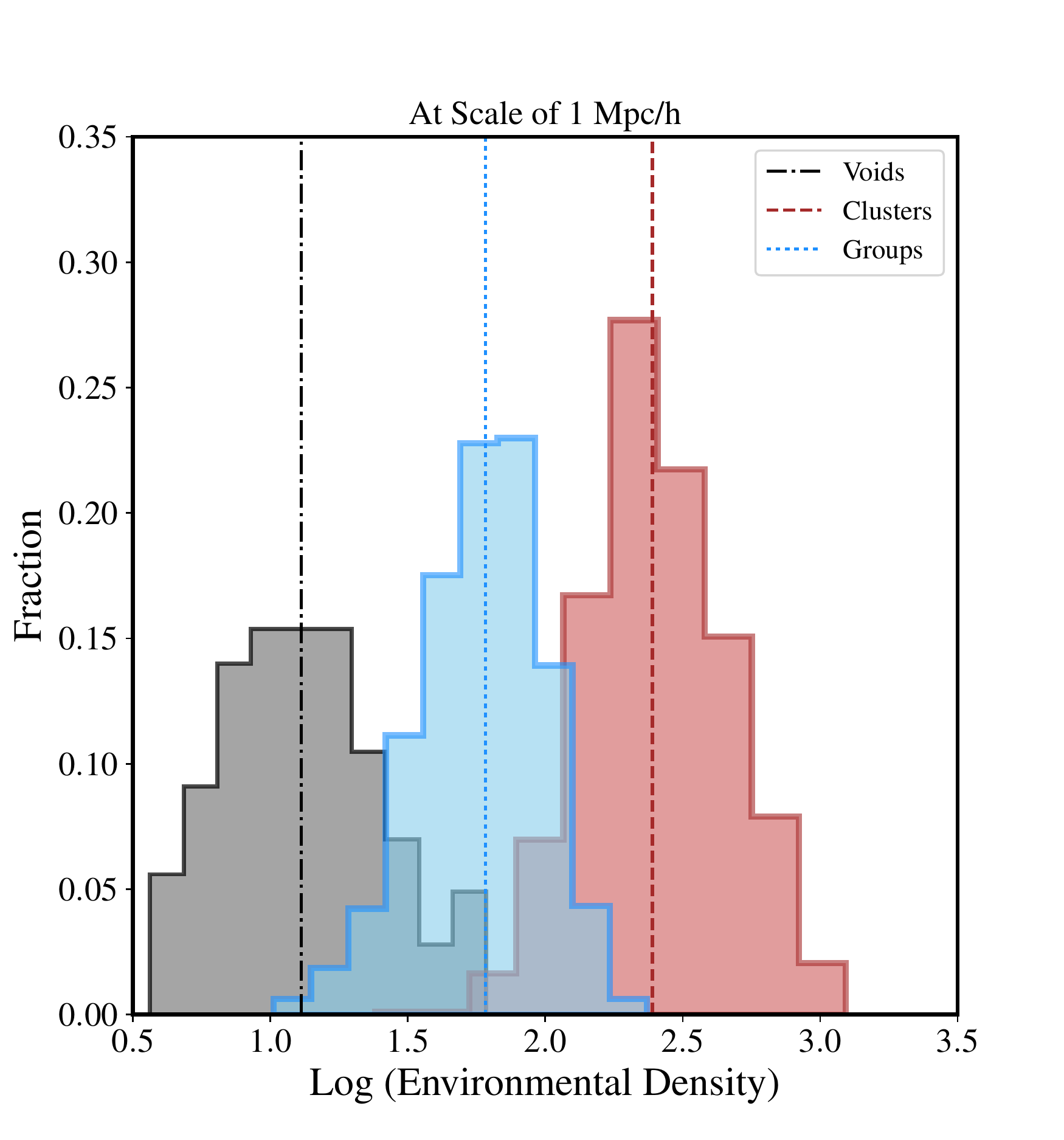}
\vspace{3. mm}
\centering
\includegraphics[width=0.4\textwidth]{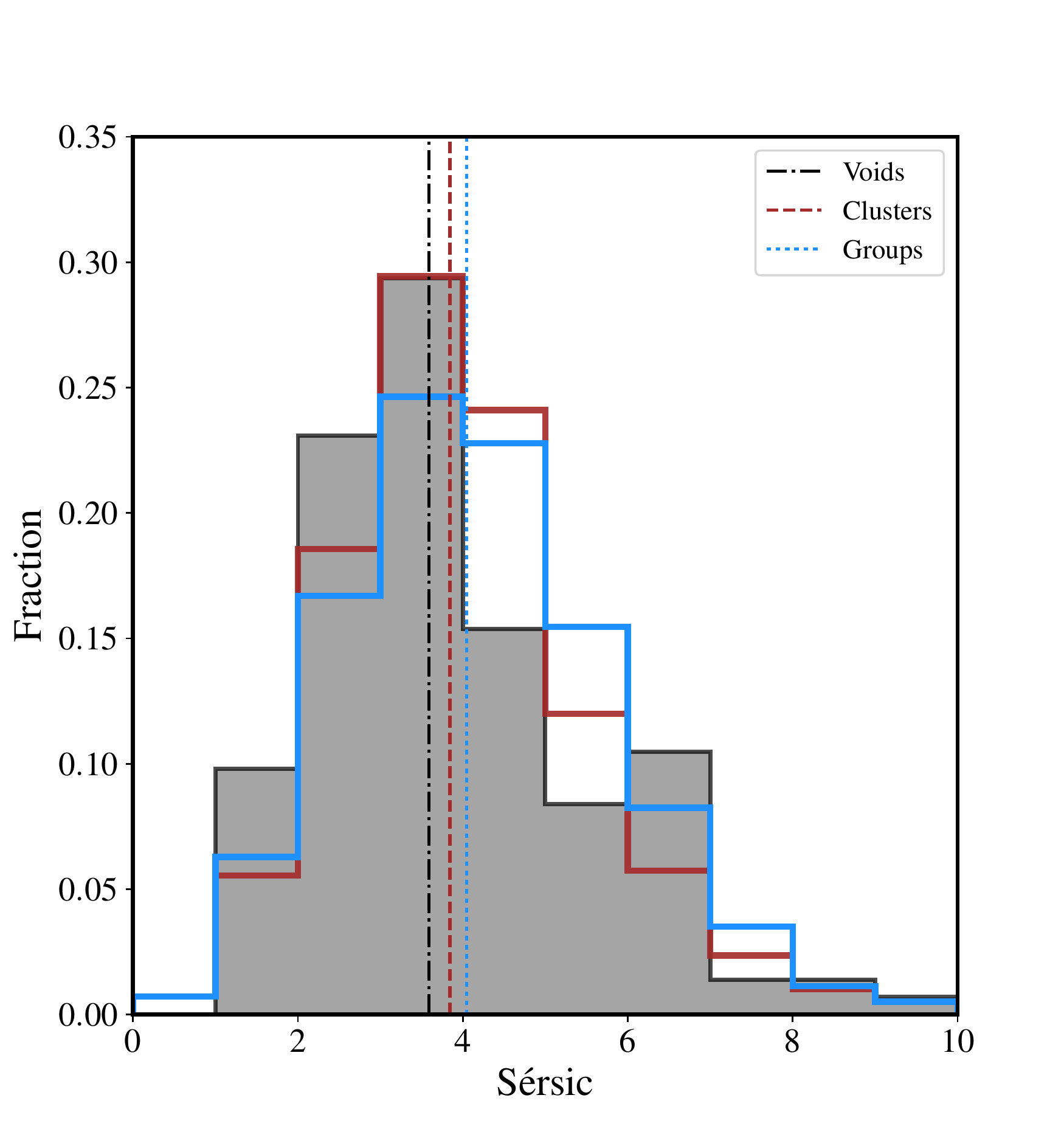}
\caption{\textit{Top Panel}: The distribution of the normalized environmental density at the scale of 1 $Mpc/h$ of isolated quiescent void galaxies (gray histogram) is compared to the ones for the quiescent galaxies in groups (blue histogram) and clusters (red histogram). The vertical lines depict the median values for each sample.  \textit{Bottom Panel}: Histograms of the distributions of the \ser indices of the quiescent galaxies for the samples probing different environments.} 
\label{fig2}
\end{figure}

The bottom left panel of Figure \ref{fig1}, shows the distribution of these galaxies in the  $g-r$ color versus $M_{r}$ plane. The inset histogram illustrates the distributions of the $u-g$ color of these quiescent galaxies (red ones) compared to the general distributions of $u-g$ color of all galaxies in voids. This indicates again that the quiescent galaxies of our sample are indeed red, and their redness is not due to the effects of the dust. The histograms in the right bottom panel of Figure \ref{fig1} illustrates the distributions of their $r$-band \ser indices ($n$) (details of their measurements are explained in Section 3). The majority (90\%) of the quiescent galaxies have high \ser indices ($n \gtrsim 2$). 

To study the role of environment on the properties of the quiescent galaxies, we compare our isolated isolated quiescent void galaxies to those in over-dense regions. In particular, we selected two different samples of quiescent galaxies from the group and cluster catalog of \cite{Tempel2014}, based on the same $urz$ color-color criteria. The first sample contains quiescent galaxies in clusters (high density regions), which have more than 15 spectroscopic members, and the second sample includes quiescent galaxies in groups (intermediate density regions), which have between 4 to 8 spectroscopic members. All galaxies are brighter than -19 in the $r-band$ filter and lie in the redshift range of $0.01 < z < 0.05$ as the original void galaxy sample. The final group/cluster samples contain $7\times$ more objects in comparison to the isolated quiescent ones in voids (i.e., 1029 objects for each sample of the groups and clusters). However, their galaxy stellar mass distribution is the same.  The normalized environmental density of galaxies for the smoothing scale of 1 Mpc/h from \cite{Tempel2014} is shown in the top panel of Figure \ref{fig2}, and the vertical lines in this plot representing the medians of environmental density for each sample. As expected, cluster galaxies reside in a significantly different environmental density than isolated void galaxies. We note that we only used color-color criteria for selection quiescent galaxies and we have not applied any additional morphological constrain. However, as  can be seen from the bottom panel of Figure \ref{fig2}, the quiescent samples from different environments follow a similar \ser index distribution with consistent median values.\\


\begin{figure*}
\centering
\includegraphics[width=\textwidth]{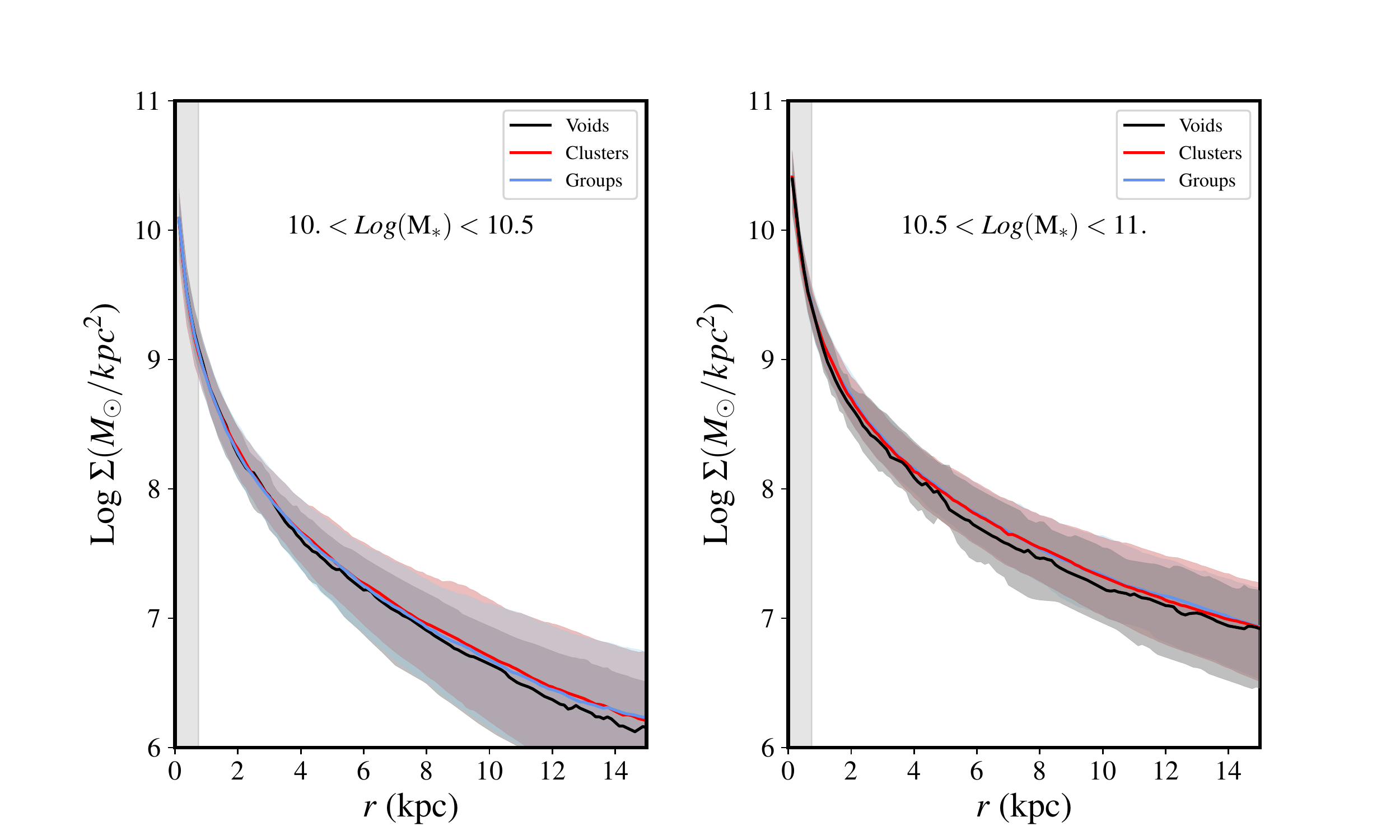}
\caption{Comparison between median surface stellar mass density of the isolated quiescent galaxies in voids (black lines) to the ones from their counterparts in the groups (blue lines) and clusters (red lines) split into two stellar mass bins of $10^{10.} - 10^{10.5} \msun$ (left panel) and $10^{10.5} - 10^{11.} \msun$ (right panel). The quiescent galaxies in groups and clusters have slightly higher stellar mass surface densities profiles at their outer regions compared to the isolated galaxies. The shaded regions indicate $1-\sigma$ scatter of the stellar mass profiles.}
\label{fig3}
\end{figure*}

\begin{figure}
\includegraphics[width=0.5\textwidth]{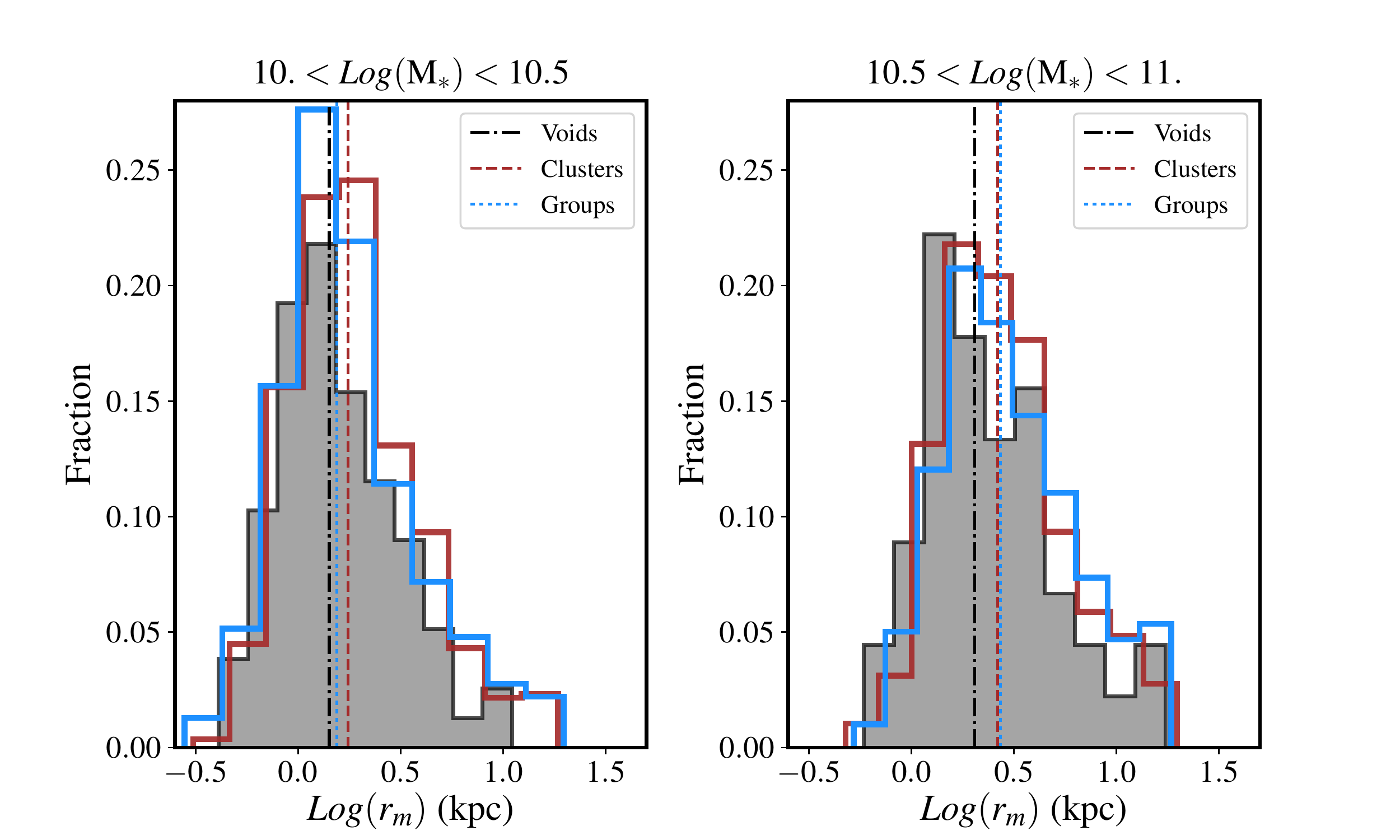}
\vspace{5. mm}
\includegraphics[width=0.5\textwidth]{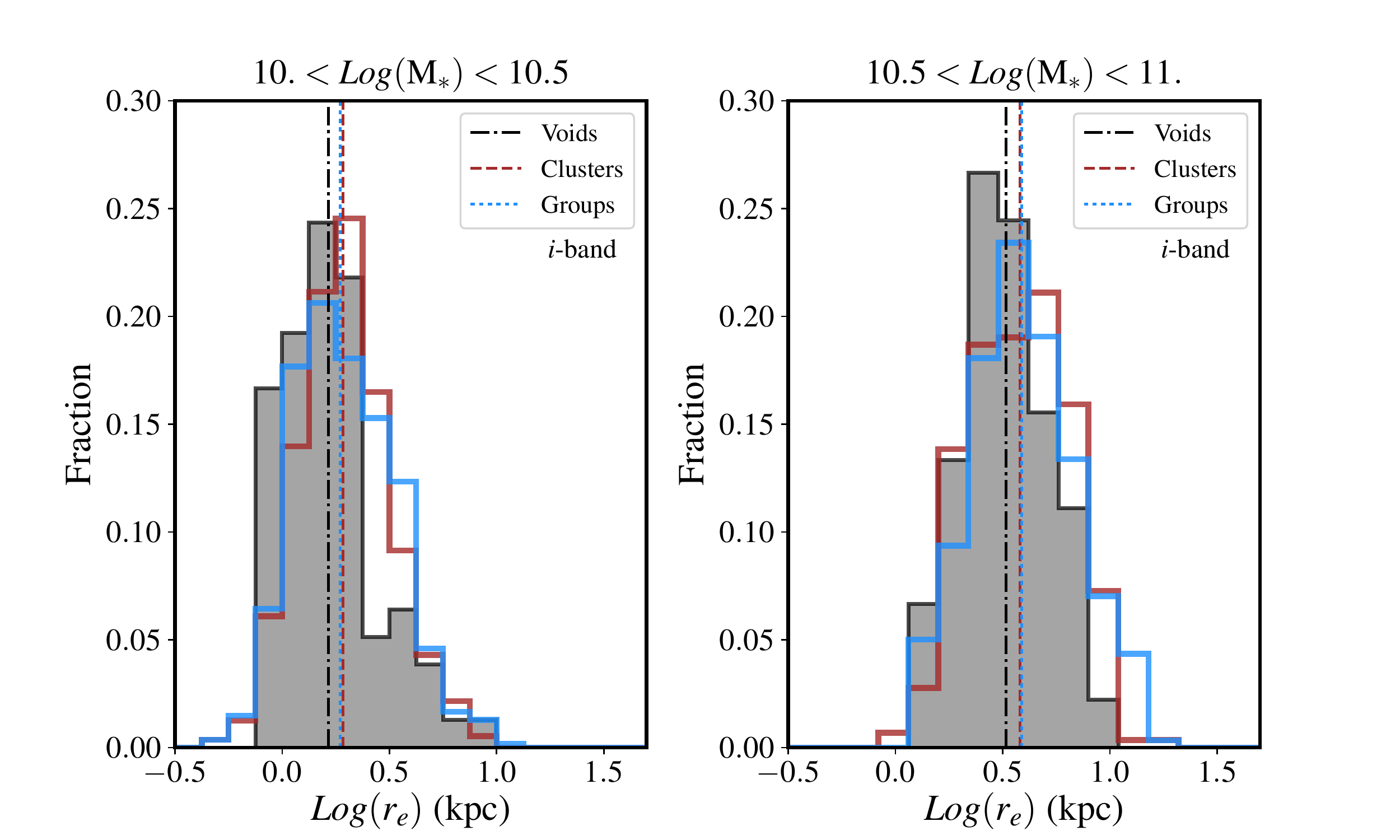}
\caption{\textit{Top panels}: Histograms showing the distributions of the half-mass sizes of isolated quiescent galaxies in voids (gray area) in comparison with quiescent galaxies in groups (blue lines) and clusters (red lines) in two different stellar mass bins. The dotted and dashed lines depict the median of the half-mass sizes for each sample. \textit{Bottom panels}: The same as top panels but for the half-light sizes in the $i$-band.  This shows that at fixed mass, isolated quiescent galaxies in voids have smaller half-mass/half-light sizes compared to their counterparts in high galaxy density environments. }
\label{fig4}
\end{figure}

\section{Methodology}
\label{method}
For this study, we derive the light and stellar mass profiles of all galaxies in our samples. The methodology for deriving the radial light and mass profiles are described in details in \cite{mosleh2017}. In brief, we use $u, g, r, i, z$  images of the galaxies from the SDSS Data Release 7 (DR7) \citep{abazajian2009} with additional Galactic extinction corrections described in \cite{Schlafly2011}. We obtain the best-fit single \ser model \citep{sersic1963, sersic1968, graham2005} of each galaxy in different filters,  using GALFIT v3 \cite{peng2010} \citep[see also][for the fitting procedure]{mosleh2013}. For each object, the $r$-band image is used as a reference to derive the geometrical properties, including their central positions, ellipticities and position angles. These properties from the $r$-band are used while fitting the single \ser profiles of the galaxies in the other filters. This ensures that systematics due to the mismatched centers and orientations are suppressed. The final light-profiles are then circularized to remove the effect of the ellipticity \citep{mosleh2011, mosleh2012}. 

The point spread function (PSF)-corrected light profiles are then converted to the stellar mass profiles. For this, the best-fit spectral energy distributions (SED) are found at each radius. As described in \cite{mosleh2017}, we divide the light profiles into small bins (0.25 kpc) and use the signal-to-noise of the $g-r$ color for adaptive binning of the light profiles (in order to increase the signal-to-noise ratio at larger radii),  prior to the SED fitting. Then, we use iSEDfit code \citep{moustakas2013} to find the best SED model at each radial bin. Following \cite{kaufmann2003a}, we use the \cite{BC2003} stellar population models, assuming the \cite{chabrier2003} initial mass function (IMF) and allowing a random star-burst on top of the exponential declining star-formation history. The stellar mass profiles are used for deriving the half-mass radii. For this, the outer parts of the stellar mass-density profiles are extrapolated using single \ser models. The half-mass radii are then found by integrating the mass-profiles out to 100 kpc.

It is important to emphasize that to remove the effect of mass loss, the stellar mass profiles in this study are corrected to the total mass (i.e., integrated star-formation history, including stellar masses, remnants, and mass which is returned back into the interstellar medium). We also note that about $\sim3 - 5\%$ of the objects in the samples are excluded from the analysis, due to their large uncertainties in their derived light profile parameters (including objects at the edge of images or close to very bright objects).\\

\begin{figure}
\includegraphics[width=0.5\textwidth]{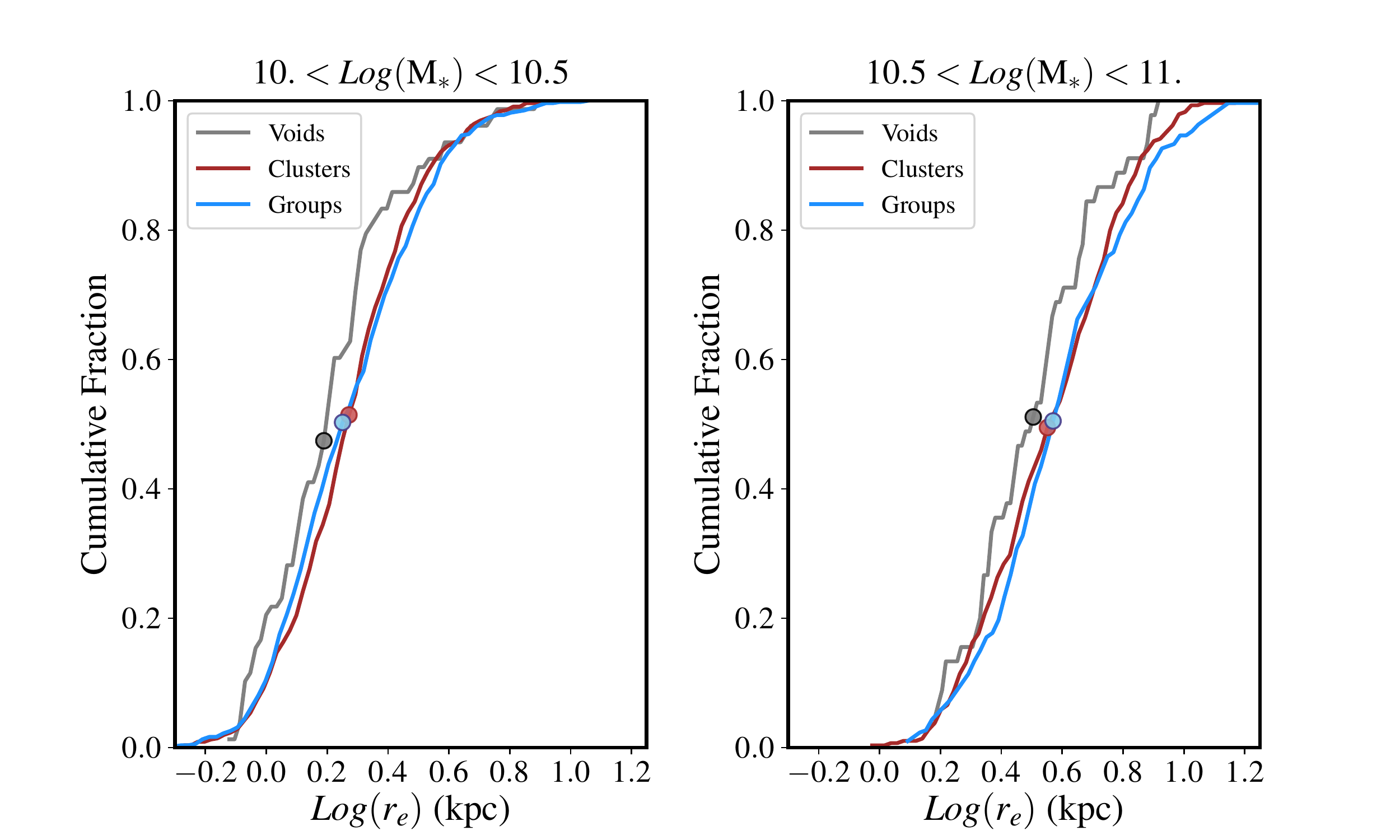}
\caption{Cumulative distribution functions of the half-light sizes in $i$-band, for the three quiescent samples in two stellar mass bins. The solid circles show the median of each sample. Again, this indicates the slight differences between sizes of quiescent galaxies in extreme over-dense and under--dense environments.}
\label{fig5}
\end{figure}

\begin{figure*}
\centering
\includegraphics[width=0.7\textwidth]{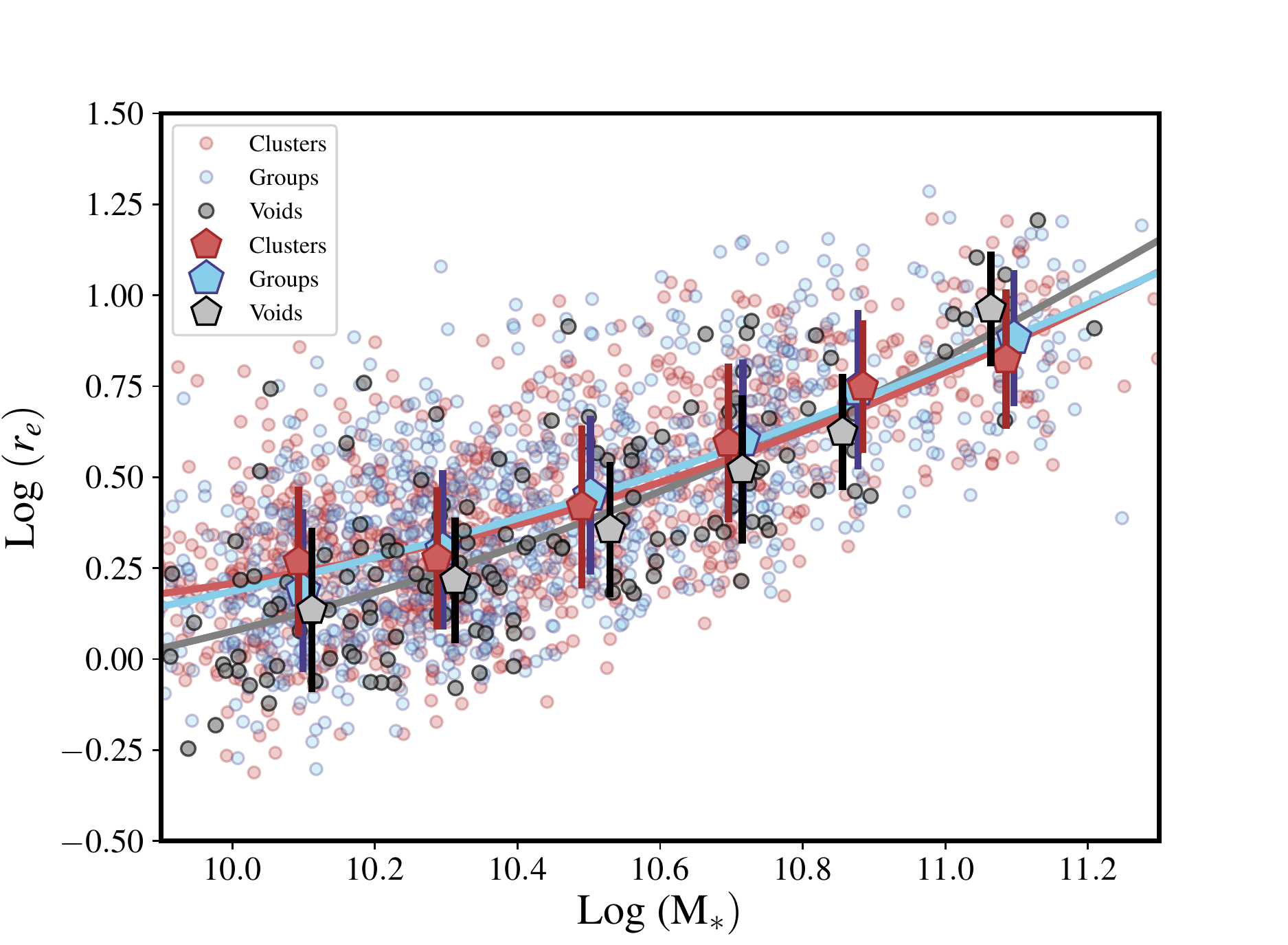}
\caption{The stellar mass-size relation for the isolated quiescent galaxies in void (gray), group (blue) and cluster (red) environments. The solid circles show the individual measurements, and the pentagons indicate the medians in each mass bin. The error bars show the $1-\sigma$ scatter of the sizes in each bin. The sizes are measured from the single-\ser best-fit models to the individual galaxy images in the $i$-band. The relation is similar for all environment but with slightly smaller sizes for the isolated quiescent galaxies in the voids for stellar masses of $10^{10} - 10^{11} \msun$. The solid lines depict their best-fit models to the data points (see text for more details).}

\label{fig6}
\end{figure*}

\begin{figure*}
\centering
\includegraphics[width=0.7\textwidth]{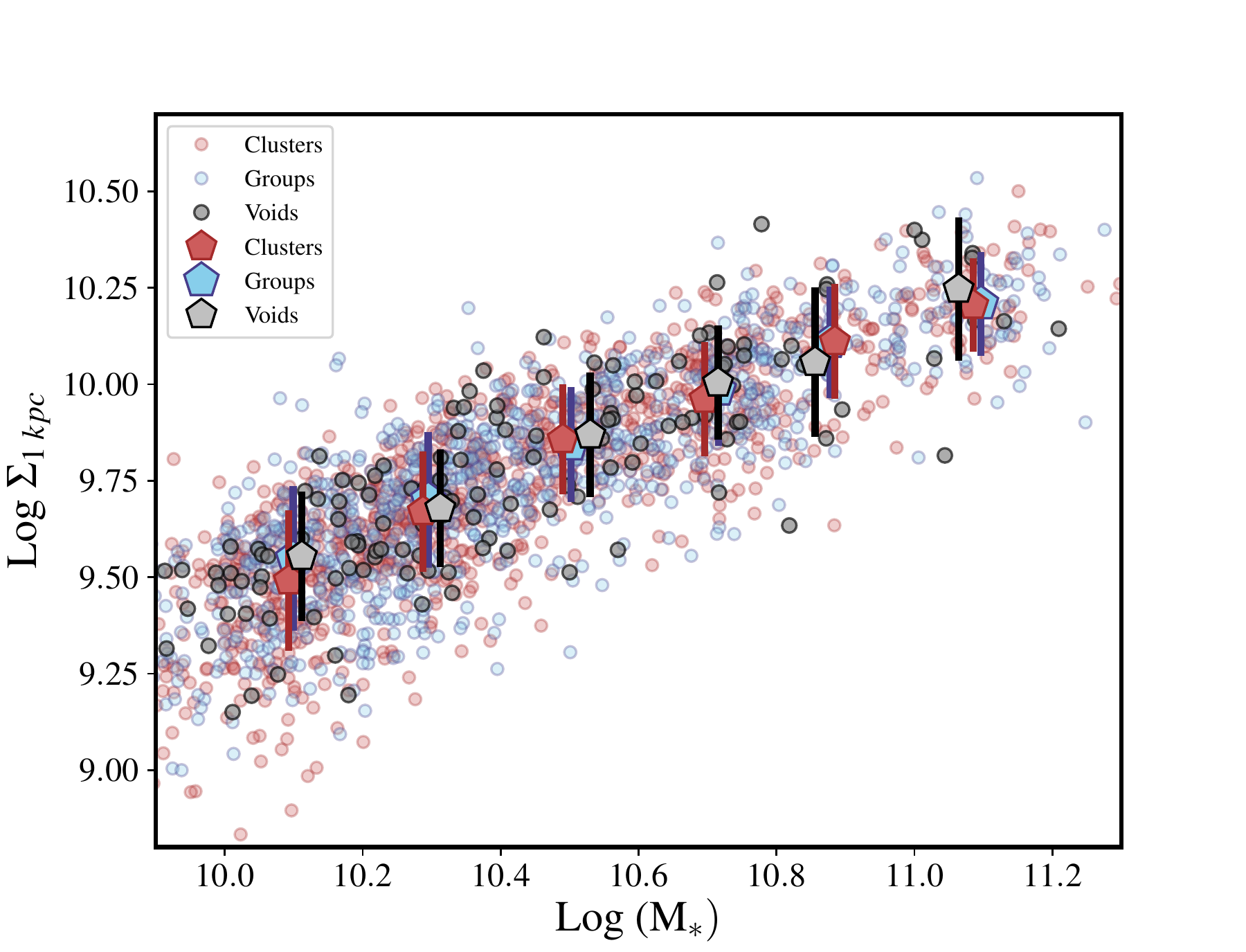}
\caption{The relation between the total stellar masses with the central mass densities ($\Sigma_1$) of quiescent galaxies in different environments. The gray, blue and red circles represent the quiescent samples of the voids, groups, and clusters and the pentagons represent the medians at each stellar mass bin. All samples follow a similar $\mstar$-$\Sigma_1$ relation in the local Universe.}
\label{fig7}
\end{figure*}

\begin{figure}
\includegraphics[width=0.48\textwidth]{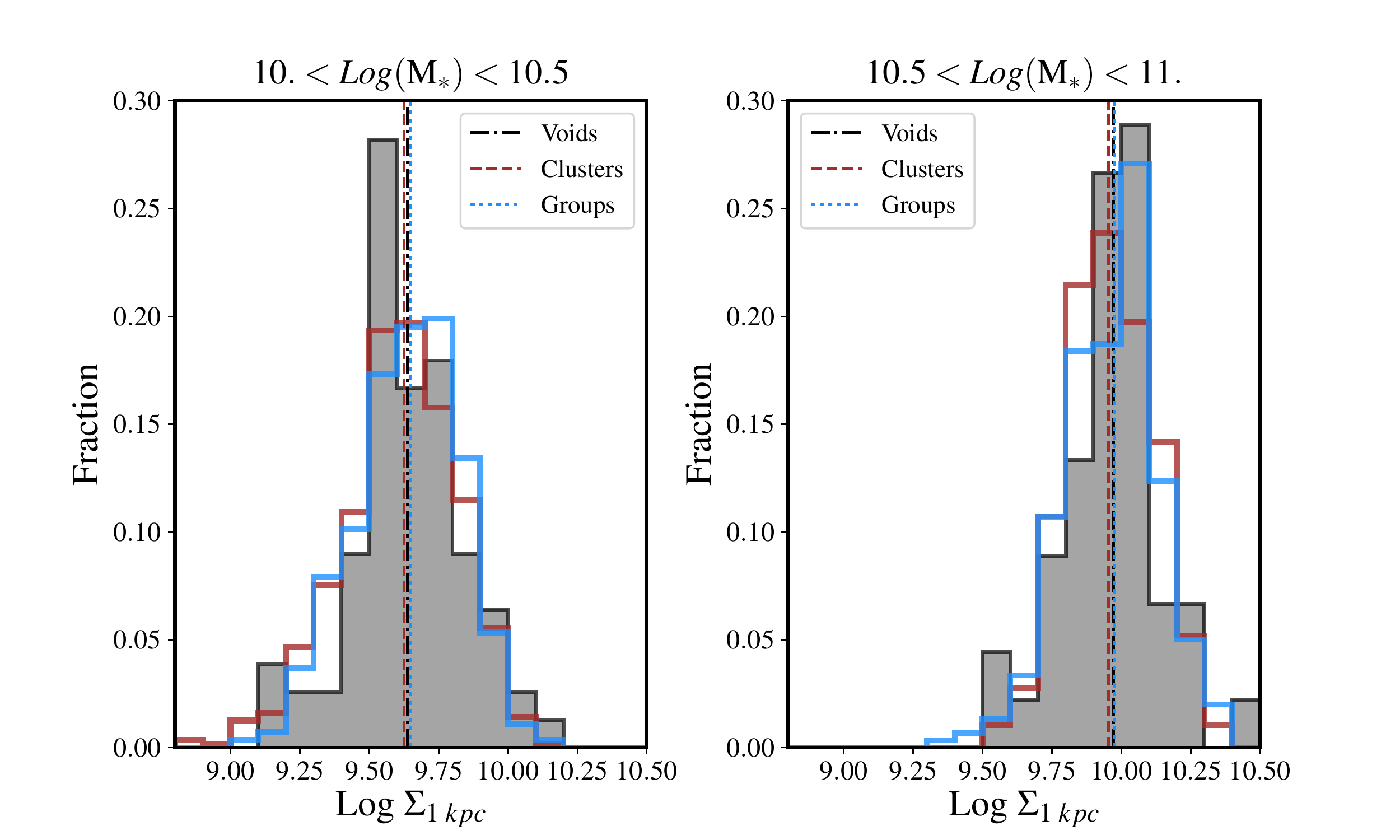}
\caption{The distributions of the central stellar mass densities ($\Sigma_1$) of the quiescent samples in different environments and at fixed stellar mass. The dotted and dashed lines are the medians for each sub-sample. At fixed masses, quiescent galaxies in the void regions have similar central mass densities to their counterparts in groups and clusters. }
\label{fig8}
\end{figure}


\section{Results}
\label{results}

As mentioned in the introduction, the stellar mass profiles of galaxies and the related properties such as their half-mass radii (or half-light radii) and their central mass densities, in addition to the kinematic parameters, are important parameters as they evolve with cosmic time and retain information about the history of their stellar mass assembly \citep[e.g.,][]{naab2009,  dokkum2010, oser2012, newman2012, tacchella2016a, Genel2017}. In this section, we compare these parameters for the samples of quiescent galaxies in different environments. 

\subsection{Mass-Size Relation}
We start by comparing the median stellar mass profiles of the samples in Figure \ref{fig3}. As the majority of the quiescent galaxies have stellar masses in the range of $10^{10}$ and $10^{11}\msun$, the samples are split into two stellar mass bins of $10. < \log(\mstar/\msun) < 10.5$ (left panel) and $10.5 < \log(\mstar/\msun) < 11.$ (right panel). The solid black lines illustrate the median mass profiles of the isolated quiescent galaxies in voids, and the red and blue ones present the median radial stellar masses of the comparison samples in clusters and groups, respectively. The shaded regions illustrate the $1-\sigma$ scatter around the median profiles. There is a hint that quiescent galaxies in the dense environments of groups and clusters have a slightly higher surface mass density at the outer parts compared to those in the very low-density environments of the voids. This is the same for low and high stellar mass bins. 

This can be investigated in detail by comparing the half-mass sizes of these galaxies at a fixed stellar mass. The histograms in the top panels of Figure \ref{fig4} depict the distributions of the half-mass radii of these galaxies in different environments and in two stellar mass bins (left and right panels). The gray, blue and red histograms are the distributions of the half-mass sizes for the quiescent galaxies in voids, groups, and clusters, respectively. The dashed-dotted, dashed and dotted vertical lines indicate the medians. The isolated voids quiescent galaxies in the high mass bin have a median half-mass size of $r_{m} = 2.04\pm^{0.26} _{0.14}$ kpc and are about $\sim25\%$ smaller than the quiescent galaxies in groups ($r_{m} = 2.72\pm^{0.23} _{0.24}$ kpc) and clusters ($r_{m} = 2.63\pm^{0.23} _{0.25}$ kpc).  This is the similar to the estimates in the lower mass bin: the isolated quiescent galaxies have $\sim20\%$ smaller half-mass sizes compared to the quiescent ones in clusters and groups (voids: $r_{m} = 1.42\pm^{0.08} _{0.11}$ kpc; groups: $r_{m} = 1.54\pm^{0.10} _{0.13}$ kpc; clusters: $r_{m} = 1.75\pm^{0.12} _{0.13}$ kpc). 

The bottom panels of Figure \ref{fig4}, illustrate the half-light sizes in the $i$-band, similar to the top panels. Again, the half-light sizes of the isolated void quiescent galaxies are slightly ($\sim 16\%$) smaller at fixed masses, compared to their counterparts in clusters and groups. In the stellar mass range of $10.5 < \log(\mstar/\msun) < 11. $, the median half-light sizes for the isolated quiescent galaxies is $r_{e} = 3.28\pm^{0.24} _{0.38}$ kpc, compared to the $r_{e} = 3.88\pm^{0.26} _{0.24}$ kpc for the quiescent objects in groups and $r_{e} = 3.80\pm^{0.37} _{0.34}$ kpc for clusters. 

In Figure \ref{fig5}, the cumulative distribution functions of the half-light sizes are shown for each sample in two different stellar mass bins. The circles indicate the median of sizes for each sample. For the massive bin, the $p$-value from the 2D KS test between sizes of the galaxies in clusters and groups is $0.35$, however, this value between sizes of quiescent galaxies in clusters and voids is $0.13$. This states that we can not reject the null hypothesis that the size distribution of both the void and cluster galaxies and also the group and cluster galaxies originates from the same distribution.

Moreover, the stellar mass-size (half-light sizes in the $i$-band) relation of the quiescent samples is shown in Figure \ref{fig6}. The circles represent the mass-size distributions for individual objects (gray, blue and red symbols for the isolated void, group and cluster galaxies, respectively) and the pentagons depict the median of sizes in each stellar mass bin. The error bars are the $1-\sigma$ scatter around the medians. The solid lines are representing the best-fit models to the individual data points (following Equation 4 in \cite{mosleh2013} and setting characteristic mass to $\log(M_{0}) = 10.53$). From the median points, the sizes of the quiescent galaxies in the isolated regions are slightly smaller than their counterparts in the high-density environments, in particular, for low mass galaxies. There is a hint that at the stellar masses above $10^{11}\msun$ the isolated quiescent void galaxies have larger sizes compared to the other comparison samples, however, the significance of this deviation is below 1-$\sigma$. In addition, as discussed in \cite{mosleh2013} and \cite{bernardi2014}, using single \ser models for galaxies in the nearby Universe can introduce a systematic error in size measurements of the galaxies, in particular for massive ones. As we have shown in the top panel of Figure \ref{figA} (Appendix), using the residual corrected method \citep{szomoru2010}, i.e., taking into account the residuals and deviations from a single \ser component fits, the mass-size relation of isolated quiescent in voids have no significant difference at the highest stellar mass bin, compared to their counterparts in groups and clusters. \\


\begin{figure}
\includegraphics[width=0.48\textwidth]{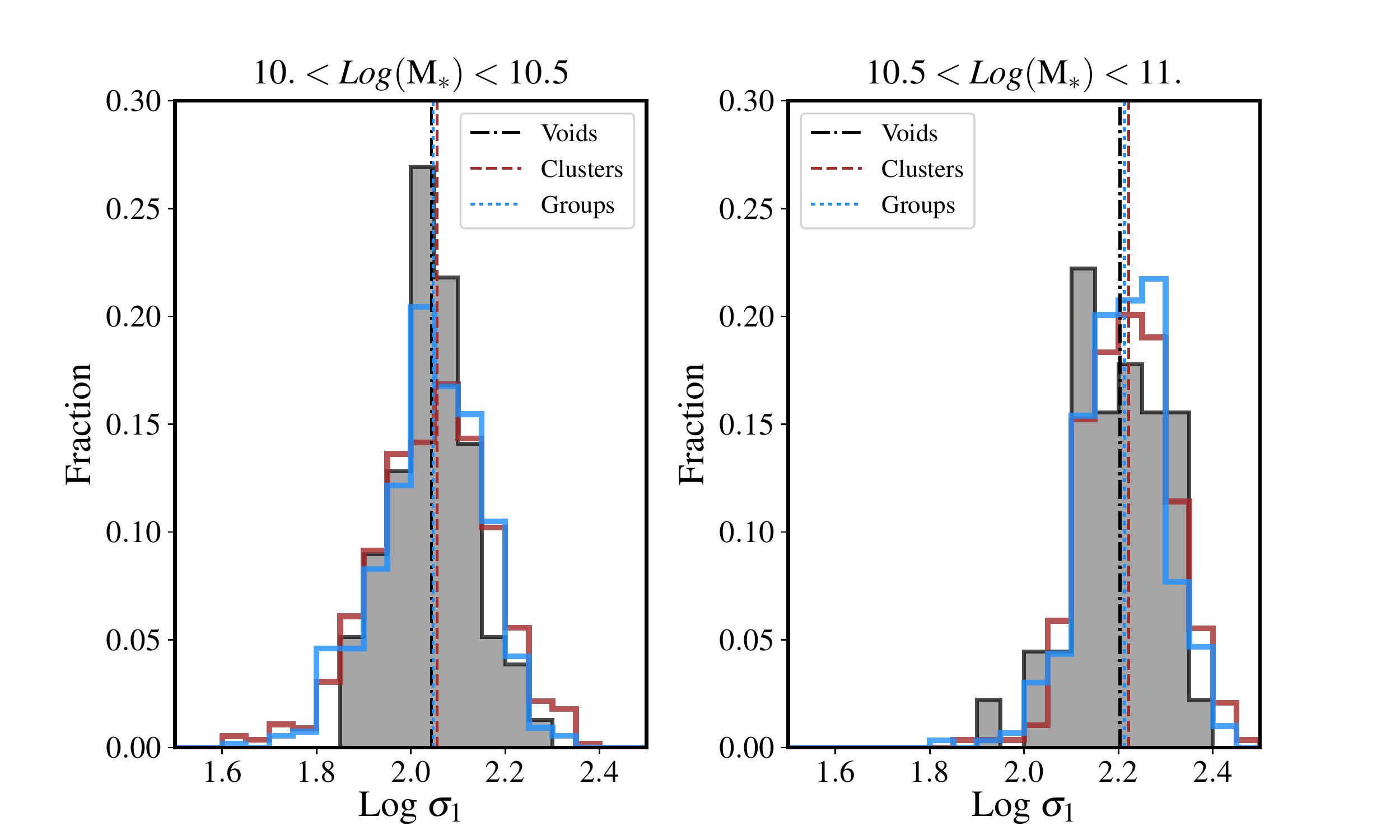}
\vspace{5. mm}
\includegraphics[width=0.48\textwidth]{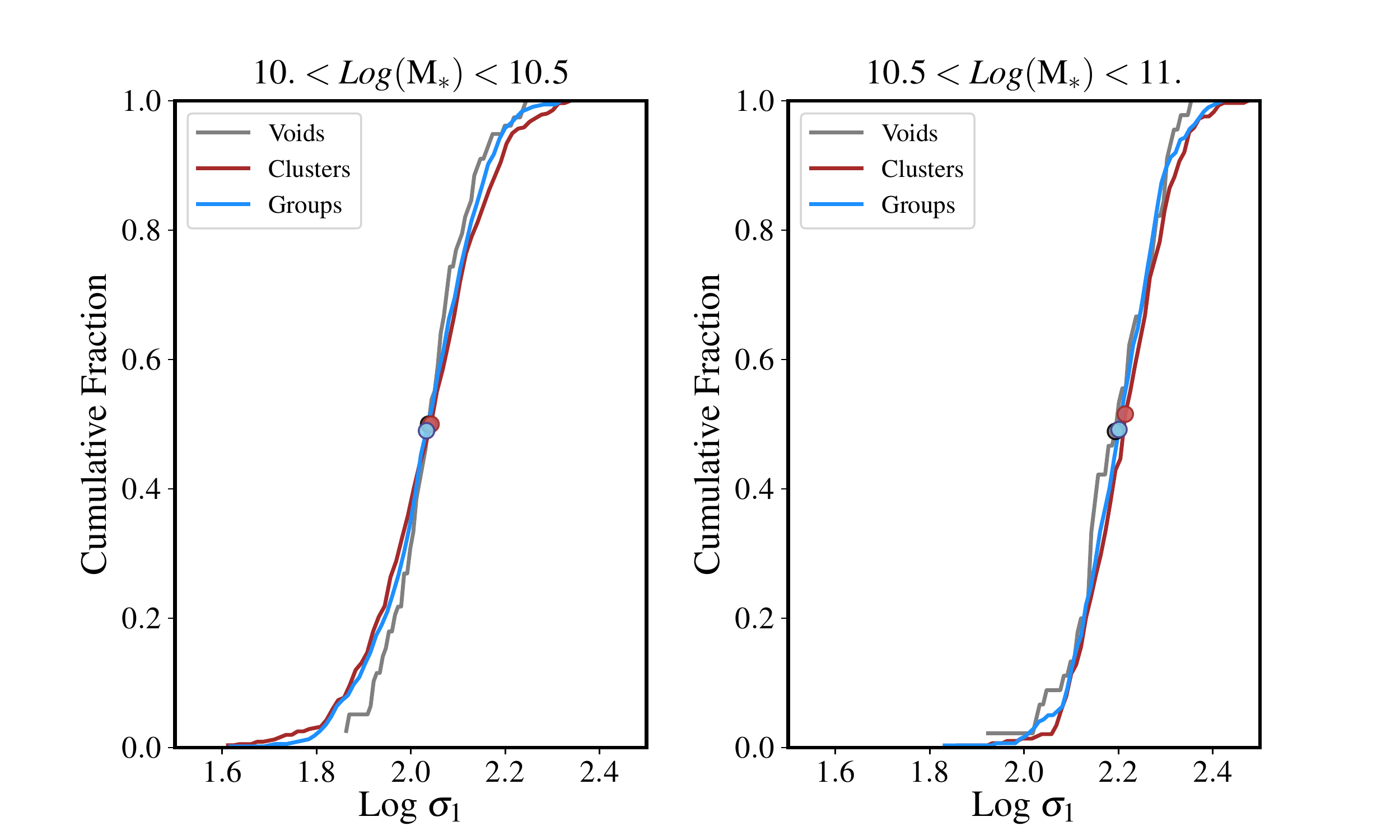}
\caption{\textit{Top panels}: Histograms showing the distributions of the central ($1$ kpc) velocity dispersion ($\sigma_1$) of the quiescent galaxies in voids (gray shaded region) in comparison with quiescent galaxies in groups (blue histogram) and clusters (red histogram) at two different stellar mass bins. The dotted and dashed lines depict the median of ($\sigma_1$) for each sample. \textit{Bottom panels}:  Cumulative distribution functions of the $\sigma_1$, for the three quiescent samples at two different stellar mass bins. The solid circles show the median of each sample.  This indicates that at least for the stellar mass range of $10 \lesssim \log(\mstar/\msun) \lesssim 11.$, there is not any significant difference between the central velocity dispersions of the quiescent galaxies in different environments.}
\label{fig9}
\end{figure}

\subsection{Central Densities \& Velocity Dispersions}

In addition to the mass-size relation, we also examined the observed properties of the central regions of these galaxies. We measured the stellar mass surface densities within an aperture of 1 kpc radius ($\Sigma_\mathrm{1 kpc}$) as a proxy for the central stellar mass density of galaxies, as used by \cite{cheung2012, saracco2012, fang2013, tacchella2015, mosleh2017}. In Figure \ref{fig7}, the correlation between the total stellar masses and the central densities ($\Sigma_1$) are shown. As shown by the median values (the pentagon symbols), the $\mstar$-$\Sigma_1$ relation for the quiescent galaxies is independent of  environment for the stellar mass range of the $10^{9.8}-10^{11.2}\msun$.  We note that the integrated star-formation history is used for measuring the central stellar masses, hence explaining any systematic in the zero-point of the $\Sigma_1$ compared to other studies. 

In Figure \ref{fig8}, the histograms of the central stellar mass densities are shown for two stellar mass bins, similar to Figure \ref{fig4}. The dotted and dashed lines depict the medians of $\Sigma_1$ in each stellar mass bin. It can be seen from this figure that there are no differences in the central stellar mass densities of the quiescent galaxies in different environments for our studied mass range. 

This can be further examined by comparing the central velocity dispersion of the galaxies (i.e., $\sigma_1$ within a circular aperture radius of 1 kpc ) at fixed masses. For this, we use the observed velocity dispersions ($\sigma_{ap}$) from the SDSS within $3''$ fiber radius ($R_{ap}$) and correct them to the velocity dispersion ($\sigma_{1}$) within 1 kpc radius  ($R_{1}$) via a relation introduced by \cite{cappellari2006}:  

\begin{equation}
(\sigma_{ap}/\sigma_{1}) = (R_{ap}/R_{1})^{-0.066}
\end{equation}

The comparison between $\sigma_{1}$ of galaxies can evaluate the differences between the dynamical properties of the central regions of the quiescent galaxies in different environments.  However, as the central velocity dispersion is related to the central density, i.e., $\Sigma_{1} \propto \sigma_{1}$ \citep[see][]{fang2013}, it is expected to see an agreement between $\sigma_1$ of different samples.

The top panels of Figure \ref{fig9} illustrate the histogram distributions of the galaxies' $\sigma_1$ in different environments. The median values are also illustrated as dashed or dotted lines. At fixed mass (at least in the range of $10^{10}-10^{11}\msun$), the central velocity dispersion of the galaxies in different environments is similar. However, from the cumulative distributions of the $\sigma_1$ (bottom panels of Figure \ref{fig9}), there is a hint that in the clusters and groups, the quiescent galaxies in the lower stellar mass bins (left bottom panel), have a wider range of $\sigma_1$ compared to the isolated quiescent galaxies. The newly accreted low-mass quiescent galaxies to the high-density regions might be a source for this slightly broader distribution of $\sigma_1$, though this needs to be investigated in future works in more detail.

\begin{figure}
\includegraphics[width=0.48\textwidth]{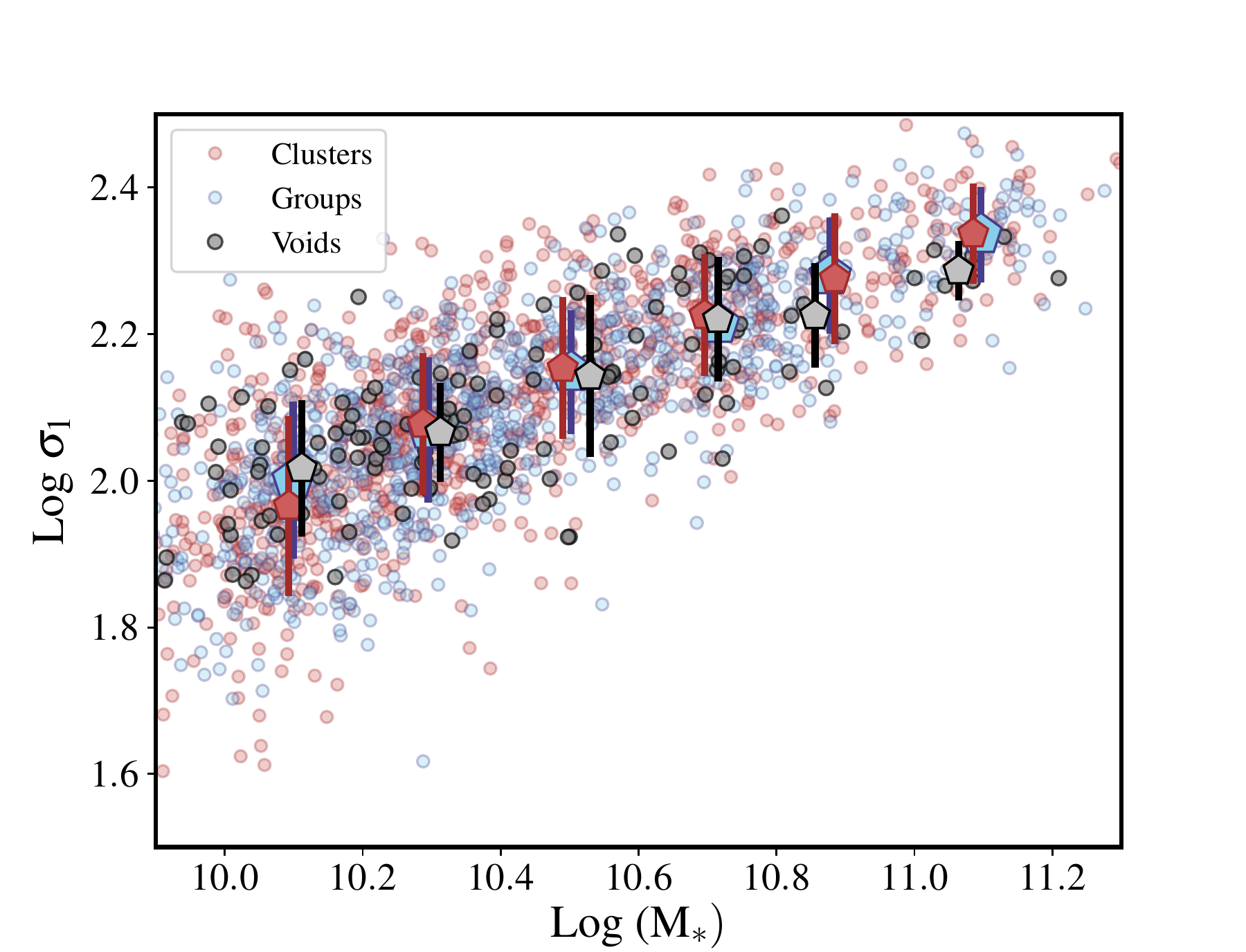}
\vspace{5. mm}
\includegraphics[width=0.48\textwidth]{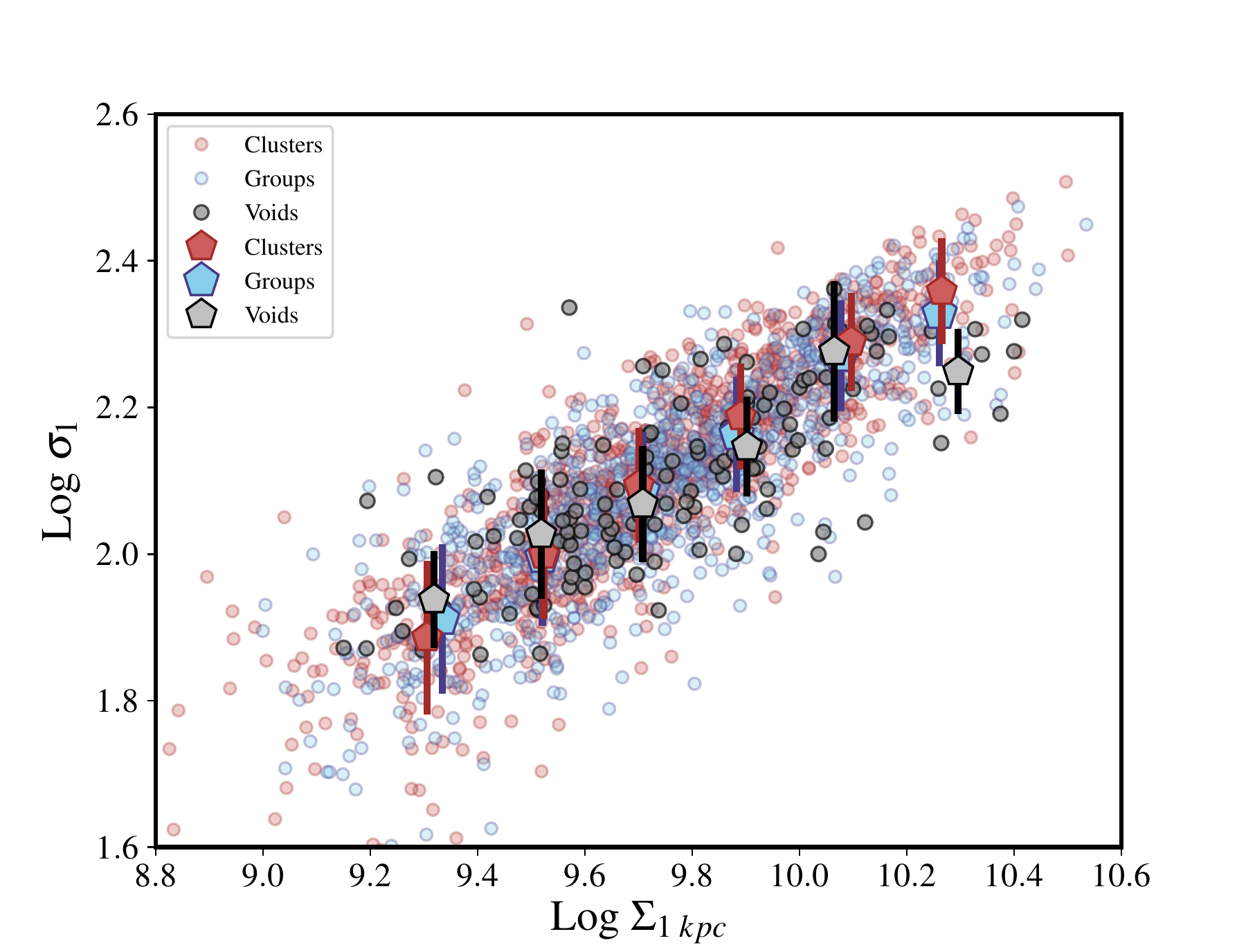}
\caption{\textit{Top panel}: The relation between the central velocity dispersion ($\sigma_1$) and the total stellar mass of the quiescent samples. \textit{Bottom Panel}: The correlation between the central mass and the central velocity dispersion ($\sigma_1$-$\Sigma_1$) of the quiescent samples. Quiescent galaxies within the stellar mass range of $\log(\mstar/\msun) \sim 10-11$ have similar $\sigma_1$-$\mstar$ and $\sigma_1$-$\Sigma_1$ relations, independent of environment. However, there is a hint that isolated quiescent galaxies with $\log(\mstar/\msun) \gtrsim 11$, have slightly lower central velocity dispersion ($\sigma_1$) compared to their counterparts in groups and clusters. }
\label{fig10}
\end{figure}

The relation between the central velocity dispersions ($\sigma_1$) with their total stellar masses of the quiescent galaxies is also consistent among samples of different environments (see top panel of Figure \ref{fig10}). However, there seems to be a tendency that in the highest stellar mass bin, i.e., $10^{11}-10^{11.2}\msun$, the isolated void quiescent galaxies have on average marginally ($\sim 12\%$) lower central velocity dispersion ($\sigma_1$) compared to their counterparts in high-density environments. This is reflected more prominently in the $\sigma_1$-$\Sigma_1$ relation, where the isolated quiescent galaxies have lower $\sigma_1$ in the highest central density $\Sigma_1$ bin.  This trend cannot be due to the \ser component fitting, as the residual corrected profiles (right panel of Figure \ref{figA} in the Appendix) also present the same behavior. If the central velocity dispersion has a correlation with the dark matter halo \citep[see e.g.,][]{zahid2016}, then the deviation at high masses could be related to the difference in their dark matter content. Nevertheless, it should be noted that the small number statistics makes it hard for a solid conclusion. \\


\section{Discussion}

In this study, we explored the properties of the isolated quiescent galaxies in the void environments. Previous works have already shown that void galaxies tend to be bluer and have a statistically higher fraction of the late-type galaxies compared to the average galaxy environments \citep[e.g.,][]{rojas2004, rojas2005, Hoyle2012, Kreckel2012,  Liu2015, beygu2017}. We have shown that the total fraction of the quiescent galaxies in the void regions is about $\sim 18\%$ down to a limiting magnitude of -19 in the $r$-band, and the fraction of the isolated quiescent ones is $\sim14\%$. This is compatible with the results of \citet{Hernandez-Toledo2010}, who found that $14.5\%$ of the isolated galaxies in the local Universe are early-types. The quenched void galaxies have the stellar mass range of $\sim 10^{10}-10^{11}\msun$.  We note that \citet{Penny2015} found that the quenched galaxies in both the void and non-void regions usually have stellar masses of more than $10^{10}\msun$, which could be a threshold for mass quenching for this volume-limited sample. Massive galaxies ($\gtrsim 10^{11.2}\msun$) are almost absent in voids and very isolated environments \citep[see Figure \ref{fig1} and also][]{Marcum2004, Varela2004, tavasoli2015}. The lower probability of (major) merger events compared to the high-density environments can be an explanation for the lack of very massive objects in voids. The curved mass-size relation for the quiescent galaxies above $\mstar \sim 2-3 \times 10^{11}$  and the lack of these massive galaxies in void regions can be a clue for the relative absences of the major mergers events \citep[e.g.,][]{Bernardi2011}. 

\citet{tavasoli2015} divided void regions into sparse and populous voids (i.e., voids with low and high number densities/or density contrasts). We examine the distribution of the isolated void quiescent galaxies and did not find any significant trend between the stellar mass and the density contrast of these galaxies.

As the environmental effects on quenching  and structural transformation are alleviated in void regions, the formation of the quiescent galaxies and their average morphological properties in these environments is intriguing, especially, once comparing them to their counterparts at high-density regions \citep[e.g.,][]{deCarvalho1992, Stocke2004, Collobert2006}. 
 
\subsection{Stellar Mass Assembly History}

In order to better understand how the isolated quiescent galaxies have assembled their stellar masses, we studied their structural properties and compared them with counterpart samples from the high-density environments.

The results of Section \ref{results} illustrate that the quiescent galaxies within the stellar mass range of $10^{9.8}-10^{11.2}\msun$ in all environments have similar properties in their central regions. Their central stellar mass densities ($\Sigma_1$) and their central velocity dispersions ($\sigma_1$) occupy a similar distribution. The isolated quiescent galaxies follow a similar $\mstar$-$\Sigma_1$ relation with respect to the same correlation for the quiescent in the groups and clusters. A similar study by \citet{Hau2006} (but for isolated ``elliptical'' galaxies) has also shown similar kinematic properties between these galaxies in the isolated and dense environments. Hence, any mechanism in the formation of the central regions of these galaxies should be the same for all environments. We should emphasize again, that this trend is true only for the intermediate stellar mass ranges. There is an indication (from the central velocity dispersion) that above this stellar mass range ($\gtrsim 10^{11.2}\msun$), galaxies in the groups and clusters might have slightly higher central velocity dispersions compared to the voids ($\sim 12\%$), though this needs a larger sample for a firm conclusion. In spite of that, the role of environment is not notable in shaping the central regions of these galaxies. There is also a hint that less massive quiescent galaxies in the groups and clusters have slightly a broader range of the central velocity dispersion ($\sigma_1$), and this could be due to a different accretion history of galaxies into these over-dense environments.  

Comparing the stellar mass profiles of these galaxies in their outer parts illustrates that isolated quiescent galaxies with stellar masses of $10^{10} - 10^{11} \msun$ in voids are about $\sim 10-25\%$ smaller than the quiescent galaxies in high-density environments. The mass-size relation of galaxies in different environments is studied by many groups at low and high redshifts.  In the local Universe (low-$z$), \citet{Reda2004, maltby2010, Nair2010, Cappellari2013ApJ, huertas2013, lorenzo2013, shankar2014, Zhao2015, lacerna2016} did not find any environmental dependence on the size of early-type or elliptical galaxies at a given mass. \citet{yoon2017} report that the early-type galaxies with stellar masses $> 10^{11.2}\msun$ in the high-density environment are 20-40 \% larger than the under-dense regions and a negligible dependence in the range of $10.7 < \log(\mstar/\msun) <11.2$. In a similar stellar mass range to this work, \citet{cebria2014} found the opposite, i.e., galaxies in the field are slightly larger than those in the high-density environments. \citet{poggianti2013a} also came to a similar conclusion and ascribed this to the age of the galaxies: older galaxies are smaller and there are overall more old galaxies in high-density regions.

At intermediate and high redshifts, the discrepancy among authors is even worse as \citet{cooper2012, Papovich2012, Raichoor2012, Zirm2012, bassett2013, lani2013, Strazzullo2013, Delaye2014, Kuchner2017} found an environmental dependency, while some groups not \citep{Rettura2010, huertas2013a, Newman2014, allen2015, kelkar2015, Allen2016, Morishita2017, saracco2017}. 

One reason for the discrepancies is the different galaxy selection methods and hence mixture of galaxies with different stellar population properties \citep{saracco2017}. In the local Universe, \citet{cebria2014} selected galaxies based on the \ser indices, however, \citet{yoon2017} used samples based on both morphology and color selections. \citet{maltby2010} and \citet{Nair2010} used morphologically selected galaxies based on visual inspections. At fixed morphology, \citet{Park2007} showed that the sizes of galaxies are independent of environment in the local Universe. If different selection methods select a different mix of old and young galaxies, and if older galaxies are smaller as found by \citet{poggianti2013a}, then one should expect inconsistency on the environmental effects on the size-mass relation among studies.

\begin{figure}
\includegraphics[width=0.48\textwidth]{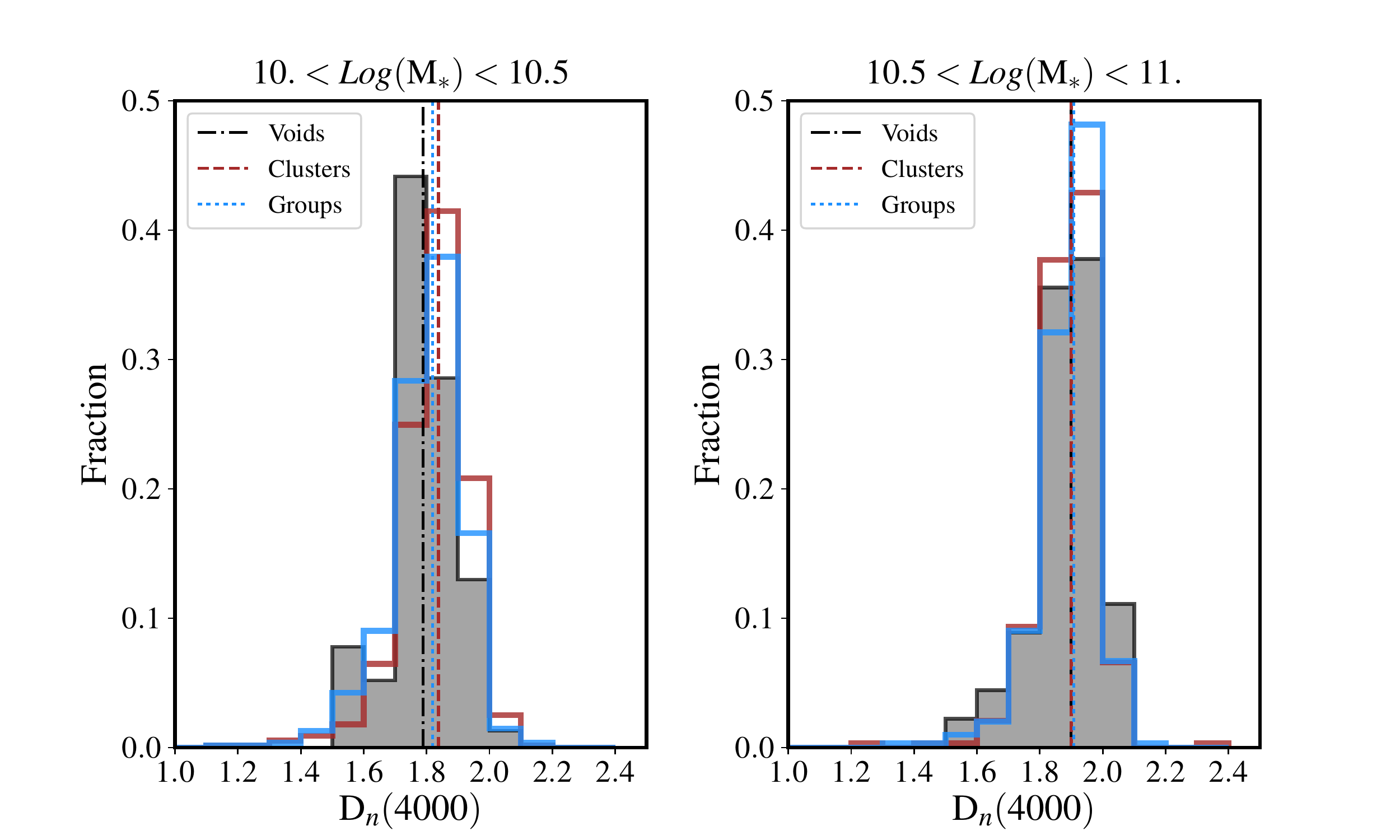}
\caption{The comparison between the age indicator of the samples, i.e., $D_n (4000)$ index at different stellar mass bins (left and right panels). The median of each sample is shown with its corresponding vertical line. At fixed stellar masses, quiescent galaxies at different environments have similar stellar population ages, but with a small difference for low mass galaxies in the voids.}
\label{fig11}
\end{figure}

\begin{figure}
\includegraphics[width=0.48\textwidth]{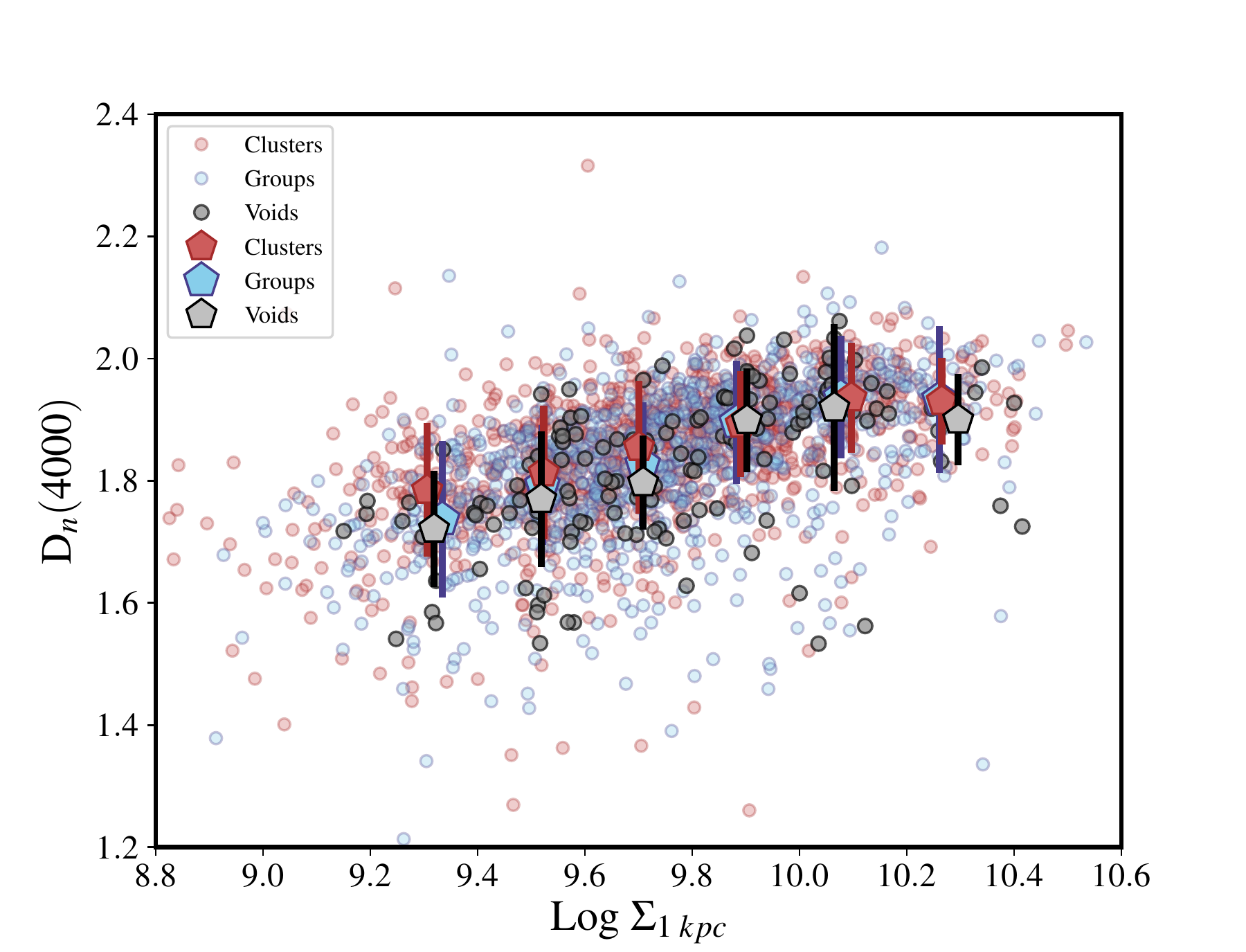}
\caption{The correlation between the stellar mass central density $\Sigma_1$ and $D_n (4000)$ index of different samples, i.e., galaxies with higher central mass densities ($\Sigma_1$) have older stellar populations. Quiescent galaxies in all environments follow a similar $\Sigma_1$-$D_n (4000)$ relation, with a hint that the low mass isolated galaxies have marginally younger average age stellar population at fixed mass.}
\label{fig12}
\end{figure}

In this work, the selection is based only on color and hence the stellar populations of galaxies. We need to investigate whether or not the differences in sizes of low and high-density regions could be due to a mixture of different galaxies in different environment. If older quiescent galaxies could have progenitors with smaller sizes \citep[e.g.,][]{Carollo2013a}, then this might contribute to the observed size differences \citep[see also][]{zahid2017}. For this purpose, we compared spectral indices of these galaxies observed by SDSS 3” fibers from MPA-JHU SDSS catalog \citep{ brinchmann2004}. As can be seen in Figure \ref{fig11}, the $D_n(4000)$ break as an age indicator of the stellar populations of the galaxies seems to have similar distributions in all environments, but with slightly smaller median value (1.78) for the lowest mass bin of the isolated void galaxies compared to the ones in the clusters (1.84). These galaxies also follow a similar $\Sigma_1$- $D_n(4000)$ relation (Figure \ref{fig12}), again with a slight hint that the isolated low mass galaxies have marginally younger stellar populations compared to the cluster members. 

\begin{figure}
\includegraphics[width=0.48\textwidth]{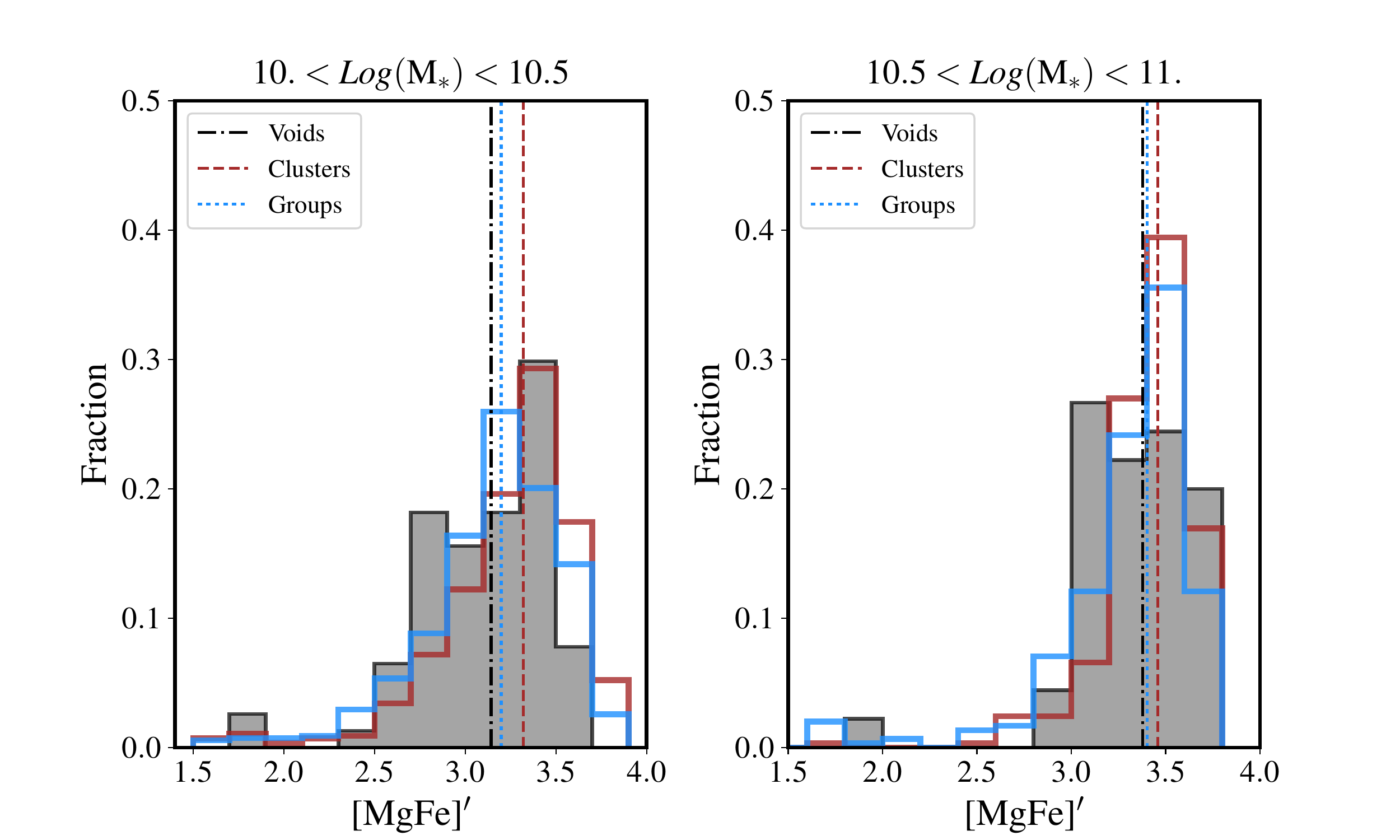}
\caption{The distributions of the stellar metallicity indicator $[MgFe]'$ of the quiescent galaxies in different environments split into two stellar mass bins (right and left panels).  The dashed and dotted lines are the medians. In the high mass bin, galaxies do not show any significant difference in their metallicity, thought in the low mass bin, quiescent galaxies in the clusters (right panel) show slightly higher stellar metallicity.}
\label{fig13}
\end{figure}

In addition, $[MgFe]'$ index is a metallicity indicator \citep[][]{Thomas2003} for diagnosing the formation history of galaxies \citep{Gallazzi2005}.  Comparing the $[MgFe]'$ of the quiescent samples (Figure \ref{fig13}) illustrates no significant differences between these populations for the massive bin. However, in the low mass bin, isolated quiescent galaxies show lower median metallicity compared to those in the clusters (This needs more investigations using detail analysis of the spectra of these galaxies and is planned for the future works.).  \citet{Denicolo2005a} also find similar velocity dispersion-index correlations for the early-type galaxies in different environments. We note that, due to the 3'' sizes of the SDSS fibers, these measurements are mostly representative of the central regions.  Nevertheless, the general similarities in the distributions of these parameters of the samples in different environments should express that the differences in the sizes (outer parts of the stellar mass profiles) of quiescent galaxies is not originated mainly due to a different mixture of galaxies in over-dense and under-dense regions \citep[see also][]{Alpaslan2015}. 

Another source of the disagreement among studies might be a different method for measuring sizes of galaxies, e.g., single \ser method as used by \citet{maltby2010} versus curve of growth as used by \citet{yoon2017}. In this work, we used a \ser model (and also residual corrected method as described in the appendix) and took into account the effect of M/L ratio gradient to estimate the stellar mass profiles. As noted in Sec. \ref{introduction}, the mass profiles are more robust than light profiles, hence removing any environmental dependence on the color-gradient of the samples which can affect the sizes. The stellar mass profiles of the galaxies, in particular, the central stellar mass densities ($\Sigma_1$; besides the \ser indices) showed that the differences of the quiescent galaxies in low and high-density regions can only be seen in the outer parts of these systems for the stellar mass range of $\log(\mstar/\msun) \sim 10 - 11$.

As discussed in the introduction, characterizing environments, and in particular, the definition of low-density regions, also varies among different works, from using halo masses \citep[e.g.,][]{huertas2013} to the fixed aperture methods \citep[e.g.,][]{cebria2014}. We have used a sample of ``isolated'' quiescent galaxies in the voids, i.e., extremely low-density regions, compared to field galaxies in previous works. This magnifies possible environmental effects.

The possible scenarios for having slightly larger sizes of the quiescent galaxies in the dense environments, could be related to the recently quenched large galaxies \citep{Kuchner2017, Morishita2017} or the mechanisms that are more frequent in the dense environments, such as minor mergers, galaxy harassment or tidal striping \citep{yoon2017}. The similarity between age and color of the galaxies in different environments (at least for the highest mass bin of $10^{10.5}-10^{11}\msun$) can reflect the marginal contributions of the biases introduced by recently accretion of large galaxies, at least for the massive ones. 

It has been shown from the observational analysis that the building up of the massive galaxies can be accelerated in the high-density environments \citep[][]{Malavasi2017, Mortlock2015} and the mechanisms such as minor mergers are more associated with the size growth of galaxies in these environments \citep{cooper2012, Lotz2013}. Hence, this effect should be better traced at high redshifts \citep{cooper2012, lani2013, Delaye2014}. If environmental effects of the mass-size relation are negligible in the local Universe, then one scenario could be the acceleration of the growth of the cluster galaxies at high redshifts and slowing down at later times, while the growth of the field galaxies accelerated at recent epochs. Consequently, this can make galaxies' mass profiles similar in field and cluster environments in the local Universe. According to this scenario, if the evolution of galaxies in void regions would be slower than in other regions, then the observed size differences compared to the high-density regions can be due to the environmental effects, and mechanisms such as minor mergers or harassment are possibly involved. Therefore, our analysis shows that the environments can affect the outer regions of the quiescent galaxies as predicted by the semi-analytic models \citep{shankar2014}. As the central velocity dispersion of different samples is similar, any environment process should mostly affect the outer regions.  In a recent study by \cite{Spindler2017}, the differences in the sizes of central and satellites at fixed central velocity dispersion are also shown and this is suggested to be related to the processes such as minor mergers. The use of extremely low and high-density environments in our study magnifies any deviations and hence the maximum differences ($10-25\%$) can be assumed as an upper limit of size differences for the intermediate stellar mass range galaxies. Studying the stellar mass profiles of a large sample of galaxies in different environments, at intermediate and high redshifts, will be the next step to examine this scenario.\\



\subsection{Quenching Mechanism}

The physical processes involved in quenching galaxies are of a matter of debate and consequently, in the isolated regions, it is crucial to understand the mechanisms for transition of galaxies into quenched ones. 

The similarities among many properties of the quenched samples with stellar masses of $\log(\mstar/\msun) \sim 10-11$ is suggestive that the role of environment is marginal compared to the cessation of their star-formation activity via physical processes related to their mass. Quiescent galaxies in all environments follow a similar $\mstar$-$\Sigma_1$ and $\mstar-\sigma_1$ relations. It can also be seen from Figure \ref{fig12} that in all environments, galaxies with higher central density have older stellar populations \cite[see also][]{tacchella2017}. These similarities between the central properties of these galaxies and the total stellar masses have been debated as the quenching mechanism might be related to the central regions of these galaxies. However, as discussed in \cite{Lilly2016}, the correlation does not prove that there is a casual link between quenching and the surface mass density. 

Studying different physical processes of quenching independent of environment is beyond the scope of this paper. In the future, we will investigate the gas fraction of these galaxies and the fraction of active galactic nuclei (AGNs) in the quiescent galaxies in different environments. \\

\section{Summary}
In this study, we investigate the properties of the isolated quiescent galaxies in the extremely low density environments, i.e., void regions, and compared them to their counterparts in groups and clusters. We found that:

\begin{itemize}
\item The quiescent galaxies in voids have intermediate stellar masses between $\sim 10^{10}-10^{11}\msun$ without any massive objects beyond $\gtrsim 10^{11.2}\msun$.
\item These galaxies only populate $\sim18\%$ of the whole void galaxies and $\sim 85\%$ of them are isolated.
\item The stellar mass central density ($\Sigma_1$) and the central velocity dispersion of the quiescent galaxies are similar at fixed masses in different environments. 
\item Isolated quiescent galaxies in the void regions are slightly ($25\%$ at most) smaller than their counterparts at the high-density environments.
\item The structures of the isolated quiescent galaxies differs very little with the cluster ones, hence suggesting a minor role of the environment in shaping the scaling relations (at least for stellar mass range of $\sim 10^{10}-10^{11}\msun$).  \\
\end{itemize}

\begin{acknowledgments}
We thank the referee for the comments that helped to improve the manuscript. Furthermore, we are grateful to Margaret Geller, Jabran Zahid and Alvio Renzini for insightful discussions.

Funding for the SDSS and SDSS-II has been provided by the Alfred P. Sloan Foundation, the Participating Institutions, the National Science Foundation, the U.S. Department of Energy, the National Aeronautics and Space Administration, the Japanese Monbukagakusho, the Max Planck Society, and the Higher Education Funding Council for England. The SDSS Web site is http://www.sdss.org/.

The SDSS is managed by the Astrophysical Research Consortium for the Participating Institutions. The Participating Institutions are the American Museum of Natural History, Astrophysical Institute Potsdam, University of Basel, University of Cambridge, Case Western Reserve University, University of Chicago, Drexel University, Fermilab, the Institute for Advanced Study, the Japan Participation Group, Johns Hopkins University, the Joint Institute for Nuclear Astrophysics, the Kavli Institute for Particle Astrophysics and Cosmology, the Korean Scientist Group, the Chinese Academy of Sciences (LAMOST), Los Alamos National Laboratory, the Max-Planck-Institute for Astronomy (MPIA), the Max-Planck-Institute for Astrophysics (MPA), New Mexico State University, Ohio State University, University of Pittsburgh, University of Portsmouth, Princeton University, the United States Naval Observatory, and the University of Washington.\\

\end{acknowledgments}

\appendix
\section{Residual Corrected Profiles}
\begin{figure}
\includegraphics[width=0.48\textwidth]{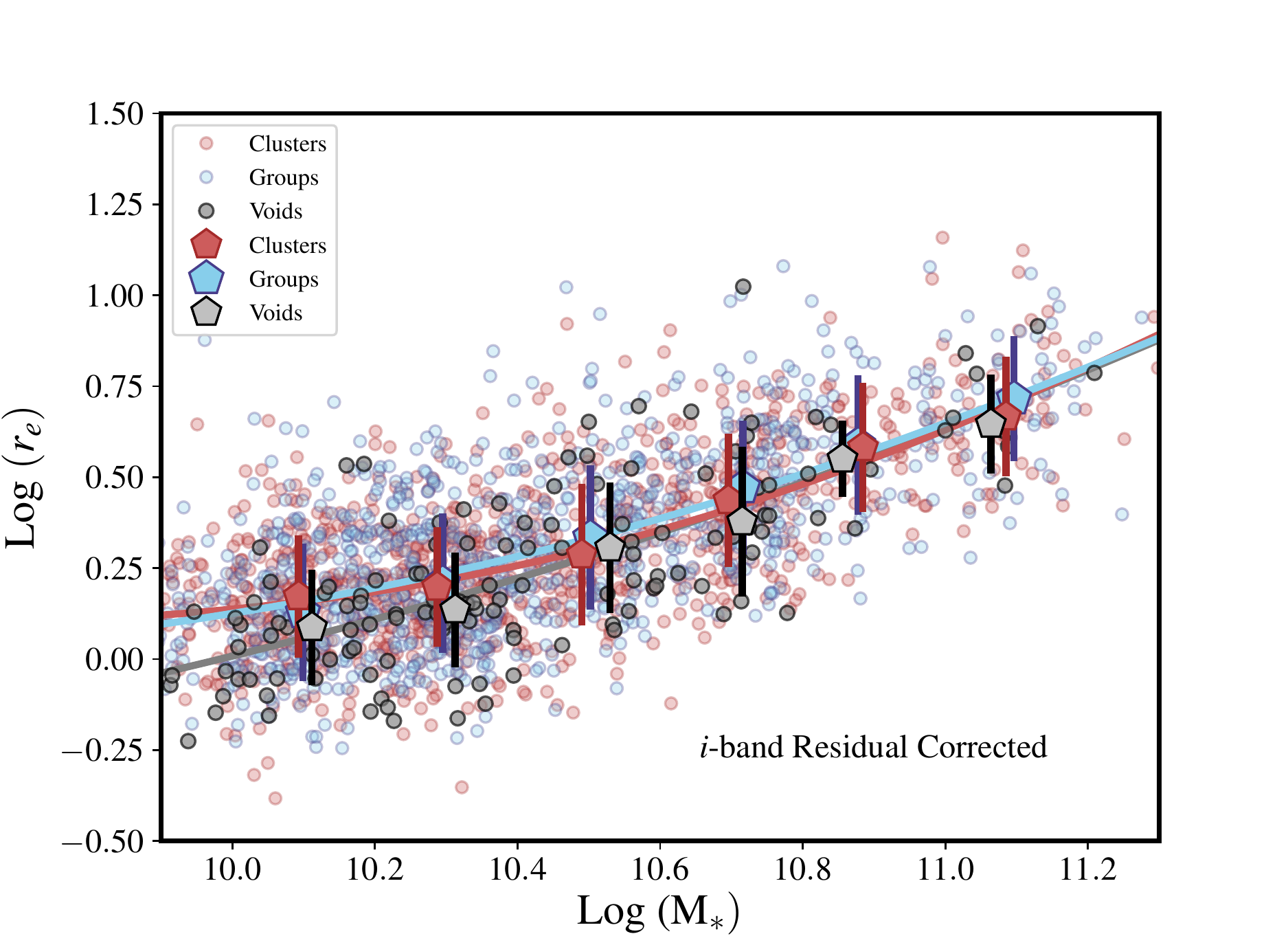}
\vspace{5 mm}
\includegraphics[width=0.48\textwidth]{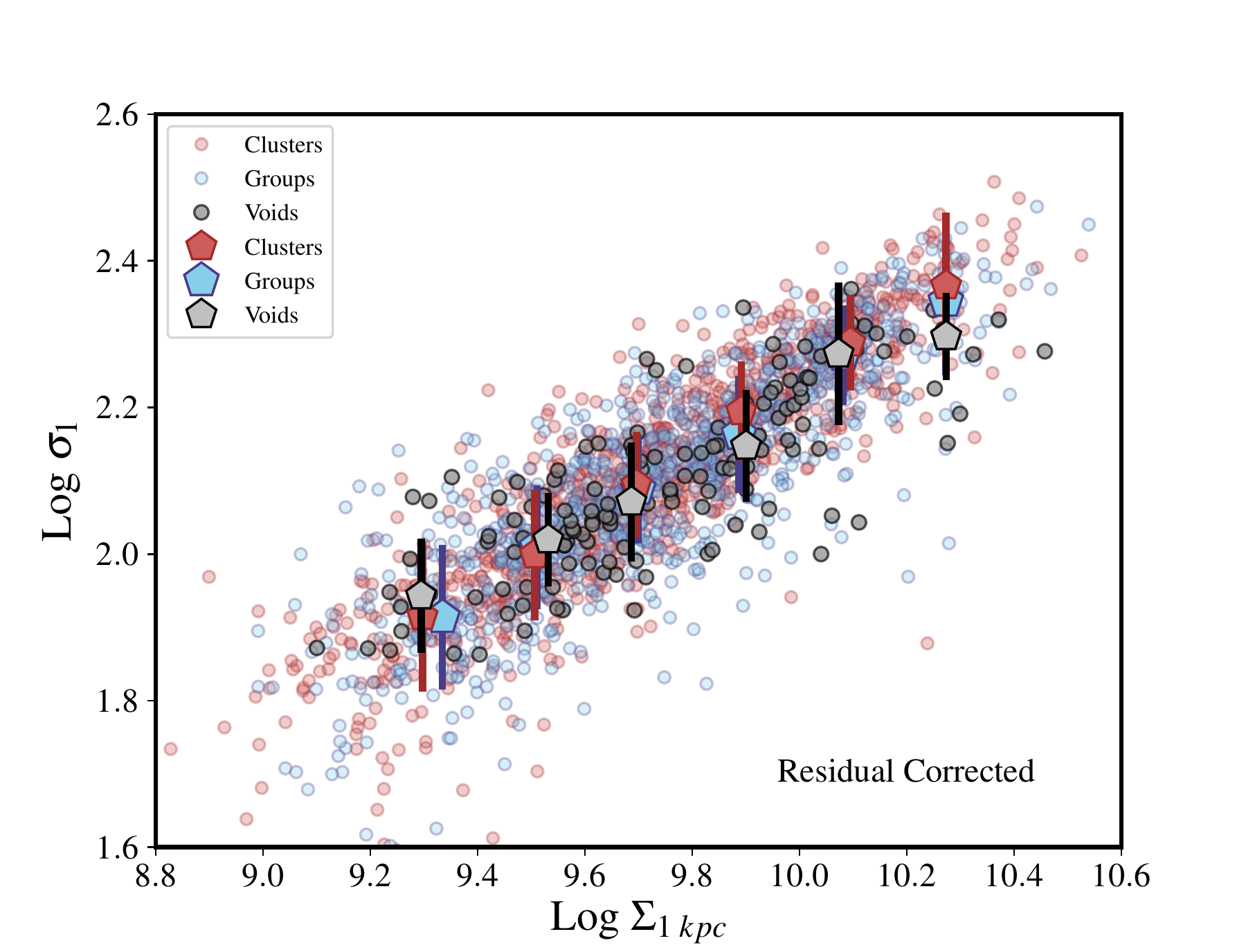}
\caption{\textit{Left panel}: The mass-size relations for the residual-corrected sizes \citep{szomoru2010} in the $i$-band for galaxies in this study. This indicates that any systematic error in size measurement or deviation from a single \ser models is not the origin of the size differences between galaxies in over-dense and under-dense environments. \textit{Right panel}: The $\sigma_1$ -$\Sigma_1$ relation for the residual corrected profiles of galaxies. Similar to the Figure \ref{fig9}, quiescent galaxies at different environments follow a similar $\sigma_1$ -$\Sigma_1$ relation, with a hint of deviation at the highest central density bins for the void quiescent galaxies.}
\label{figA}
\end{figure}

As pointed by \citep{mosleh2013, bernardi2014}, the light profiles of the nearby galaxies are better fitted with two-component \ser models and using single-component \ser profiles can introduce systematics in measuring their mass profiles. In order to address that the results of this paper do not depend on using single \ser models, we derived the stellar mass profiles of galaxies using PSF residual-corrected method \cite{szomoru2010} and therefore, overcoming any deviation from pure \ser models. \citep[see also appendix of][]{mosleh2017}. In  Figure \ref{figA},  we compared the mass-size relation and the central stellar mass density of galaxies with the total stellar masses of quiescent galaxies in three different environments the same as of Figure \ref{fig6} \& \ref{fig7}. It can be seen that the deviation of single component models does not alter the results of this paper.  \\

\end{document}